\documentclass[preprint2]{aastex62}

\usepackage{amsmath}
\usepackage{cancel}
\usepackage{float}

\shorttitle{Updated Opacities for DSEP}
\shortauthors{Boudreaux et al.}

\begin{document}

\title{Updated High-Temperature Opacities for The Dartmouth Stellar Evolution
Program and their Effect on the Jao Gap Location}

\correspondingauthor{Emily M. Boudreaux}
\email{emily.m.boudreaux.gr@dartmouth.edu,\\emily@boudreauxmail.com}

\author[0000-0002-2600-7513]{Emily M. Boudreaux}
\affiliation{Department of Physics and Astronomy, Dartmouth College, Hanover, NH 03755, USA}

\author[0000-0003-3096-4161]{Brian C. Chaboyer}
\affiliation{Department of Physics and Astronomy, Dartmouth College, Hanover, NH 03755, USA}

\received{9/20/2022}
\revised{12/12/2022}
\revised{1/23/2023}
\accepted{1/24/2023}

\begin{abstract}

	The Jao Gap, a 17 percent decrease in stellar density at M$_{G} \sim$ 10
	identified in both Gaia DR2 and EDR3 data, presents a new method to probe
	the interior structure of stars near the fully convective transition mass.
	The Gap is believed to originate from convective kissing instability
	wherein asymmetric production of $^{3}$He causes the core convective zone
	of a star to periodically expand and contract and consequently the stars’
	luminosity to vary. Modeling of the Gap has revealed a sensitivity in its
	magnitude to a population’s metallicity primarily through opacity. Thus
	far, models of the Jao Gap have relied on OPAL high-temperature radiative
	opacities. Here we present updated synthetic population models tracing the
	Gap location modeled with the Dartmouth stellar evolution code using the
	OPLIB high-temperature radiative opacities. Use of these updated opacities
	changes the predicted location of the Jao Gap by $\sim$0.05 mag as compared
	to models which use the OPAL opacities. This difference is likeley too
	small to be detectable in empirical data.

\end{abstract}


\section{INTRODUCTION}\label{sec:intro}
Due to the initial mass requirements of the molecular clouds which collapse to form
stars, star formation is strongly biased towards lower mass, later spectral
class stars when compared to higher mass stars. Partly as a result of this
bias and partly as a result of their extremely long main-sequence lifetimes,
M Dwarfs make up approximately 70 percent of all stars in the galaxy. Moreover,
some planet search campaigns have focused on M Dwarfs due to the relative ease
of detecting small planets in their habitable zones \citep[e.g.][]{Nut08}.
M Dwarfs then represent both a key component of the galactic stellar population
as well as the possible set of stars which may host habitable exoplanets.
Given this key location M Dwarfs occupy in modern astronomy it is important to
have a thorough understanding of their structure and evolution.

\citet{Jao2018} discovered a novel feature in the Gaia Data Release 2 (DR2)
$G_{BP}-G_{RP}$ color-magnitude-diagram. Around $M_{G}=10$ there is an
approximately 17 percent decrease in stellar density of the sample of stars
\citet{Jao2018} considered. Subsequently, this has become known as either the
Jao Gap, or Gaia M Dwarf Gap. Following the initial detection of the Gap in DR2
the Gap has also potentially been observed in 2MASS \citep{Skrutskie2006,
Jao2018}; however, the significance of this detection is quite weak and it
relies on the prior of the Gap's location from Gaia data. Further, the Gap is
also present in Gaia Early Data Release 3 (EDR3) \citep{Jao2021}. These EDR3
and 2MASS data sets then indicate that this feature is not a bias inherent to
DR2.

The Gap is generally attributed to convective instabilities in the cores of
stars straddling the fully convective transition mass (0.3 - 0.35 M$_{\odot}$)
\citep{Baraffe2018}. These instabilities interrupt the normal, slow, main
sequence luminosity evolution of a star and result in luminosities lower
than expected from the main sequence mass-luminosity relation \citep{Jao2020}.

The Jao Gap, inherently a feature of M Dwarf populations, provides an enticing
and unique view into the interior physics of these stars \citep{Feiden2021}.
This is especially important as, unlike more massive stars, M Dwarf seismology
is infeasible due to the short periods and extremely small
magnitudes which both radial and low-order low-degree non-radial seismic waves
are predicted to have in such low mass stars \citep{Rodriguez-Lopez2019}. The
Jao Gap therefore provides one of the only current methods to probe the
interior physics of M Dwarfs.

Despite the early success of modeling the Gap some issues remain.
\citet{Jao2020, Jao2021} identify that the Gap has a wedge shape which has not been
successful reproduced by any current modeling efforts and which implies a
somewhat unusual population composition of young, metal-poor stars. Further,
\citet{Jao2020} identify substructure, an additional over density of stars,
directly below the Gap, again a feature not yet fully captured by current
models. 

All currently published models of the Jao Gap make use of OPAL high temperature
radiative opacities. Here we investigate the effect of using the more
up-to-date OPLIB high temperature radiative opacities and whether these opacity
tables bring models more in line with observations. In Section \ref{sec:JaoGap}
we provide an overview of the physics believed to result in the Jao Gap, in
Section \ref{sec:opac} we review the differences between OPAL and OPLIB and
describe how we update DSEP to use OPLIB opacity tables. Section
\ref{sec:modeling} walks through the stellar evolution and population synthesis
modeling we perform. Finally, in Section \ref{sec:results} we present our
findings. 

%


\section{Jao Gap}\label{sec:JaoGap}
\begin{figure}
	\centering
	\includegraphics[width=0.45\textwidth]{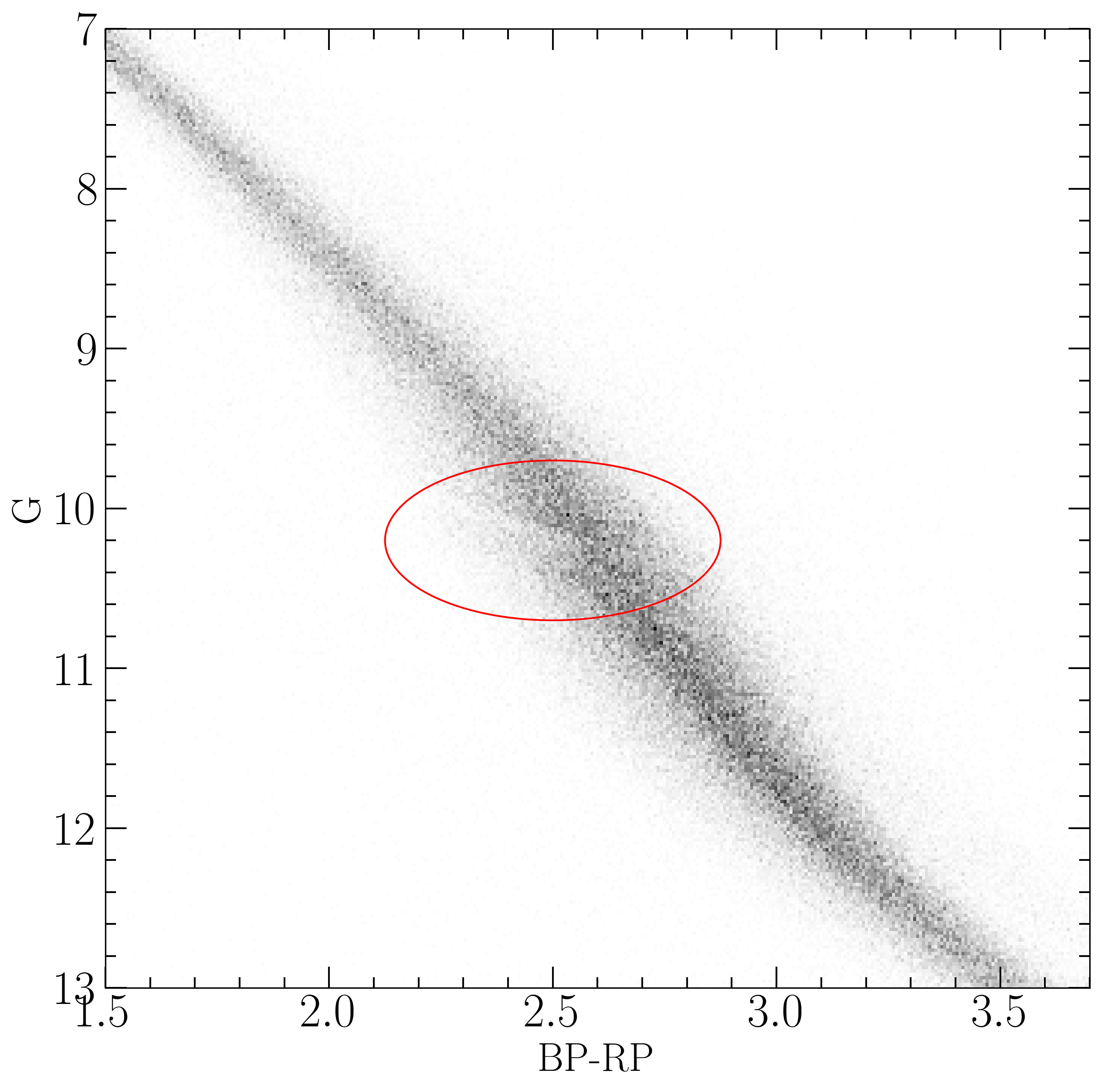}
	\caption{The Jao Gap (circled) seen in the Gaia Catalogue of Nearby Stars \citep{GaiaCollaboration2021}.}
	\label{fig:JaoGap}
\end{figure}

A theoretical explanation for the Jao Gap (Figure \ref{fig:JaoGap}) comes from
\citet{van2012}, who propose that in a star directly above the transition mass,
due to asymmetric production and destruction of $^{3}$He during the
proton-proton I chain (ppI), periodic luminosity variations can be induced.
This process is known as convective-kissing instability. Very shortly after the
zero-age main sequence such a star will briefly develop a radiative core;
however, as the core temperature exceeds $7\times 10^{6}$ K, enough energy will
be produced by the ppI chain that the core once again becomes convective. At
this point the star exists with both a convective core and envelope, in
addition to a thin, radiative layer separating the two. Subsequently,
asymmetries in ppI affect the evolution of the star's convective core.

While kissing instability has been the most widely adopted model to
explain the existence of the Jao Gap, slightly different mechanisms have also
been proposed. \citet{MacDonald2018} make use of a fully implicit stellar
evolution suite which treats convective mixing as a diffusive property.
\citeauthor{MacDonald2018} treat convective mixing this way in order to account
for a core deuterium concentration gradient proposed by \citet{Baraffe1997}.
Under this treatment the instability results only in a single mixing event ---
as opposed to periodic mixing events. Single mixing events may be more in line with
observations (see section \ref{sec:results} for more details on how periodic
mixings can effect a synthetic population) where there is only well documented
evidence of a single gap. However, recent work by \citet{Jao2021} which
identify an second under density of stars below the canonical gap, does leave
the door open for the periodic mixing events.

\begin{figure*}
	\centering
	\includegraphics[width=0.95\textwidth]{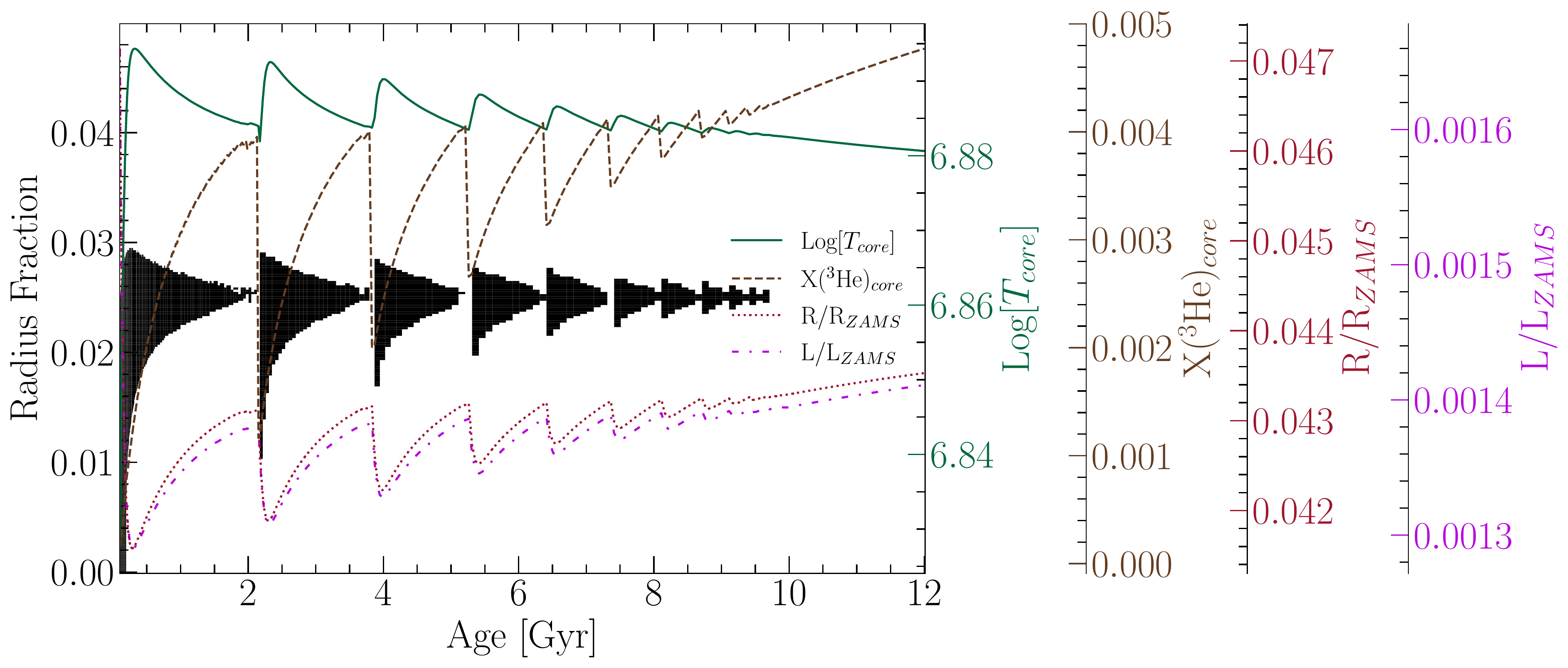}
	\caption{Diagram for a characteristic stellar model of 0.35625 $M_{\odot}$
	which is within the Jao Gap's mass range. The black shaded regions denote
	whether, at a particular model age, a radial shell within the model is
	radiative (with white meaning convective). The lines trace the models core
	temperature, core $^{3}$He mass fraction, fractional luminosity wrt. the
	zero age main sequence and fractional radius wrt. the zero age main
	sequence.}
	\label{fig:Kippenhan1}
\end{figure*}

The proton-proton I chain constitutes three reactions 
\begin{enumerate} 
	\item $p + p \longrightarrow d + e^{+} + \nu_{e}$
	\item $p + d \longrightarrow \ ^{3}\text{He} + \gamma$
	\item $^{3}\text{He} + ^{3}\text{He} \longrightarrow \ ^{3}\text{He} + 2p$ 
\end{enumerate} 
Initially, reaction 3 of ppI consumes $^{3}$He at a slower rate than it is
produced by reaction 2 and as a result, the core $^{3}$He abundance and
consequently the rate of reaction 3, increases with time. The core convective
zone expands as more of the star becomes unstable to convection. This expansion
continues until the core connects with the convective envelope. At this point
convective mixing can transport material throughout the entire star and the
high concentration of $^{3}$He rapidly diffuses outward, away from the core,
decreasing energy generation as reaction 3 slows down. Ultimately, this leads
to the convective region around the core pulling back away from the convective
envelope, leaving in place the radiative transition zone, at which point
$^{3}$He concentrations grow in the core until it once again expands to meet
the envelope.  These periodic mixing events will continue until $^{3}$He
concentrations throughout the star reach an equilibrium ultimately resulting in
a fully convective star. Figure \ref{fig:Kippenhan1} traces the evolution of a
characteristic star within the Jao Gap's mass range.

\subsection{Efforts to Model the Gap}
Since the identification of the Gap, stellar modeling has been
conducted to better constrain its location, effects, and exact cause.
Both \citet{Mansfield2021} and \citet{Feiden2021} identify that the Gap's mass
location is correlated with model metallicity --- the mass-luminosity
discontinuity in lower metallicity models being at a commensurately lower mass.
\citet{Feiden2021} suggests this dependence is due to the steep relation of
the radiative temperature gradient, $\nabla_{rad}$, on temperature and, in turn,
on stellar mass.

\begin{align}\label{eqn:radGrad}
	\nabla_{rad} \propto \frac{L\kappa}{T^{4}}
\end{align}

As metallicity decreases so does opacity, which, by Equation \ref{eqn:radGrad},
dramatically lowers the temperature at which radiation will dominate energy
transport \citep{Chabrier1997}. Since main sequence stars are virialized the
core temperature is proportional to the core density and total mass. Therefore,
if the core temperature where convective-kissing instability is expected
decreases with metallicity, so too will the mass of stars which experience such
instabilities.


The strong opacity dependence of the Jao Gap begs the question: what is
the effect of different opacity calculations on Gap properties.
As we can see above, changing opacity should affect the Gap's location in the
mass-luminosity relation and therefore in a color-magnitude diagram. Moreover,
current models of the Gap have yet to locate it precisely in the CMD
\citep{Feiden2021} with an approximate 0.16 G-magnitude difference between the
observed and modeled Gaps. Opacity provides one, as yet unexplored, parameter
which has the potential to resolve these discrepancies.

\section{Updated Opacities}\label{sec:opac}
Multiple groups have released high-temperature opacities including, the Opacity
Project \citep[OP][]{Seaton1994}, Laurence Livermore National Labs OPAL opacity
tables \citep{Iglesias1996}, and Los Alamos National Labs OPLIB opacity tables
\citep{Colgan2016}. OPAL high-temperature radiative opacity tables in
particular are very widely used by current generation isochrone grids
\citep[e.g. Dartmouth, MIST, \& StarEvol, ][]{Dotter2008,Choi2016,Amard2019}.
OPLIB opacity tables \citep{Colgan2016} are not widely used but include the
most up-to-date plasma modeling.

While the overall effect on the CMD of using OPLIB compared to OPAL tables is
small, the strong theoretical opacity dependence of the Jao Gap raises the
potential for these small effects to measurably shift the Gap's location. We
update DSEP to use high temperature opacity tables based on measurements from
Los Alamos national Labs T-1 group \citep[OPLIB,][]{Colgan2016}. The OPLIB
tables are created with ATOMIC \citep{Magee2004,Hakel2006,Fontes2016}, a modern
LTE and non-LTE opacity and plasma modeling code. These updated tables were
initially created in order to incorporate the most up to date plasma
physics at the time \citep{Bahcall2005}. 

OPLIB tables include monochromatic Rosseland mean opacities --- composed from
bound-bound, bound-free, free-free, and scattering opacities --- for elements
hydrogen through zinc over temperatures 0.5eV to 100 keV (5802 K -- 1.16$\times
10 ^{9}$ K) and for mass densities from approximately $10^{-8}$ g cm$^{-3}$ up
to approximately $10^{4}$ g cm$^{-3}$ (though the exact mass density range
varies as a function of temperature). 

\begin{figure}
	\centering
	\includegraphics[width=0.45\textwidth]{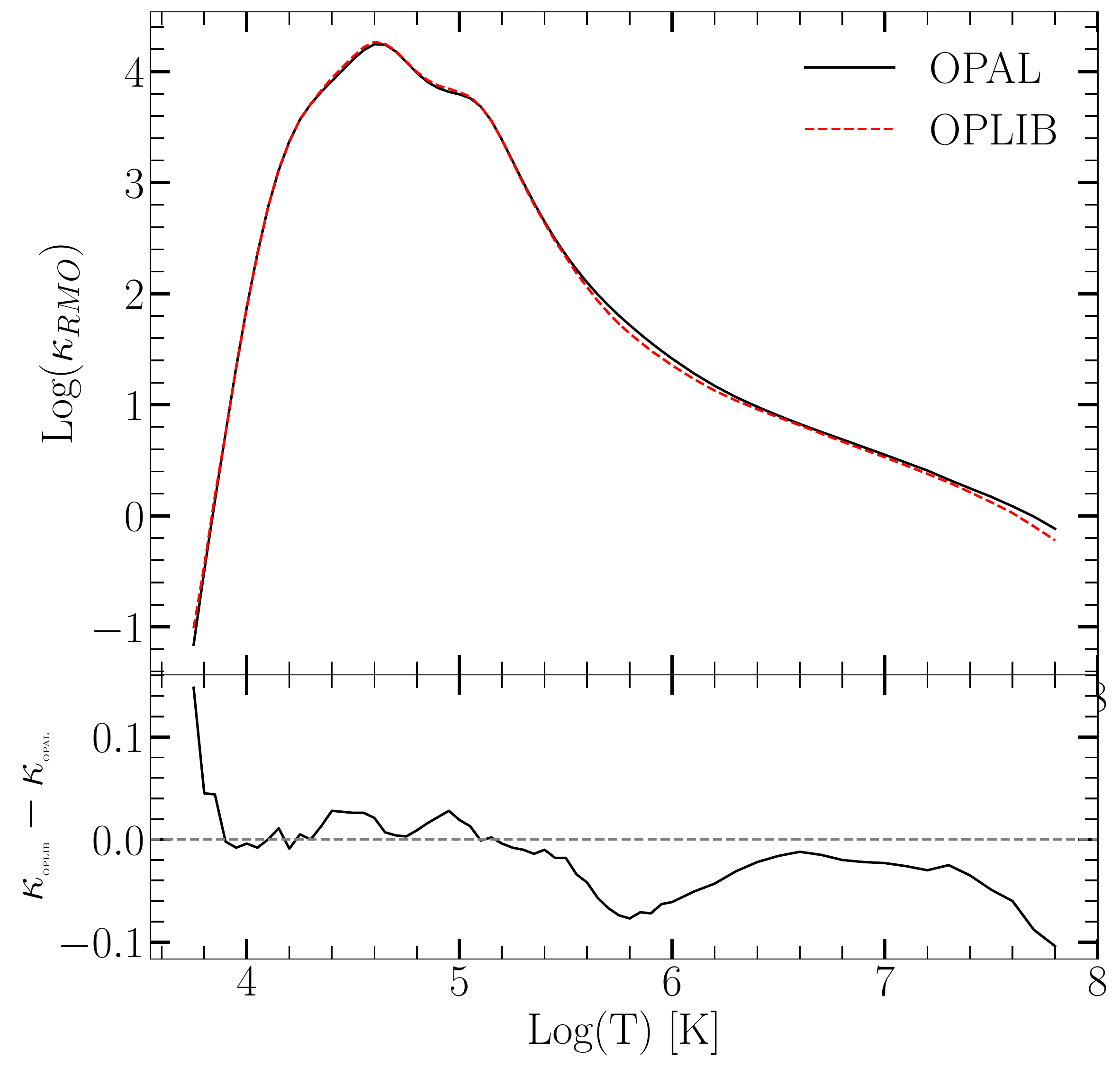}
	\caption{Rosseland mean opacity with the GS98 solar composition for both
	OPAL opacities and OPLIB opacities (top). Residuals between OPLIB opacities
	and OPAL opacities (bottom). These opacities are plotted at $\log _{10}(R)
	= -0.5$, $X=0.7$, and $Z=0.02$. $\log _{10}(R)=-0.5$ approximates
	much of the interior a 0.35 M$_{\odot}$ model. Note how the OPLIB
	opacities are systematically lower than the OPAL opacities for temperatures
	above $10^{5.2}$ K.}
	\label{fig:opacComp}
\end{figure}

DSEP ramps the \citet{Ferguson2005} low temperature opacities to high
temperature opacities tables between $10^{4.3}$ K and $10^{4.5}$ K; therefore,
only differences between high-temperature opacity sources above $10^{4.3}$ K
can effect model evolution. When comparing OPAL and OPLIB opacity tables
(Figure \ref{fig:opacComp}) we find OPLIB opacities are systematically lower
than OPAL opacities for temperatures above $10^{5}$ K. Between $10^{4.3}$ and
$10^{5} K$ OPLIB opacities are larger than OPAL opacities. These generally
lower opacities will decrease the radiative temperature gradient throughout
much of the radius of a model.

\subsection{Table Querying and Conversion}
The high-temperature opacity tables used by DSEP and most other stellar
evolution programs give Rosseland-mean opacity, $\kappa_{R}$, along three
dimensions: temperature, a density proxy $R$ (Equation \ref{eqn:R}; $T_{6} =
T\times10^{-6}$, $\rho$ is the mass density), and composition. 

\begin{align} \label{eqn:R}
	R = \frac{\rho}{T_{6}^{3}}
\end{align}

OPLIB tables may be queried from a web
interface\footnote{https://aphysics2.lanl.gov/apps/}; however, OPLIB opacities
are parametrized using mass-density and temperature instead of $R$ and
temperature. It is most efficient for us to convert these tables to the OPAL
format instead of modifying DSEP to use the OPLIB format directly. In order to
generate many tables easily and quickly we develop a web scraper
\citep[\texttt{pyTOPSScrape},][]{Boudreaux22} which can automatically retrieve
all the tables needed to build an opacity table in the OPAL format.
\texttt{pyTOPSScrape}\footnote{https://github.com/tboudreaux/pytopsscrape} has
been released under the permissive \texttt{MIT} license with the consent of the
Los Alamos T-1 group. For a detailed discussion of how the web scraper works
and how OPLIB tables are transformed into a format DSEP can use see Appendices
\ref{apx:pytopsscrape} \& \ref{apx:interp}.

\subsection{Solar Calibrated Stellar Models}\label{sec:SCSM}

In order to validate the OPLIB opacities, we generate a solar calibrated
stellar model (SCSM) using these new tables. We first manually calibrate the
surface Z/X abundance to within one part in 100 of the solar value \citep[][Z/X=0.23]{Grevesse1998}.
Subsequently, we allow both the convective mixing length parameter,
$\alpha_{ML}$, and the initial Hydrogen mass fraction, $X$, to vary
simultaneously, minimizing the difference, to within one part in $10^{5}$,
between resultant models' final radius and luminosity to those of the sun.
Finally, we confirm that the model's surface Z/X abundance is still within one
part in 100 of the solar value.

\begin{figure}
	\centering
	\includegraphics[width=0.45\textwidth]{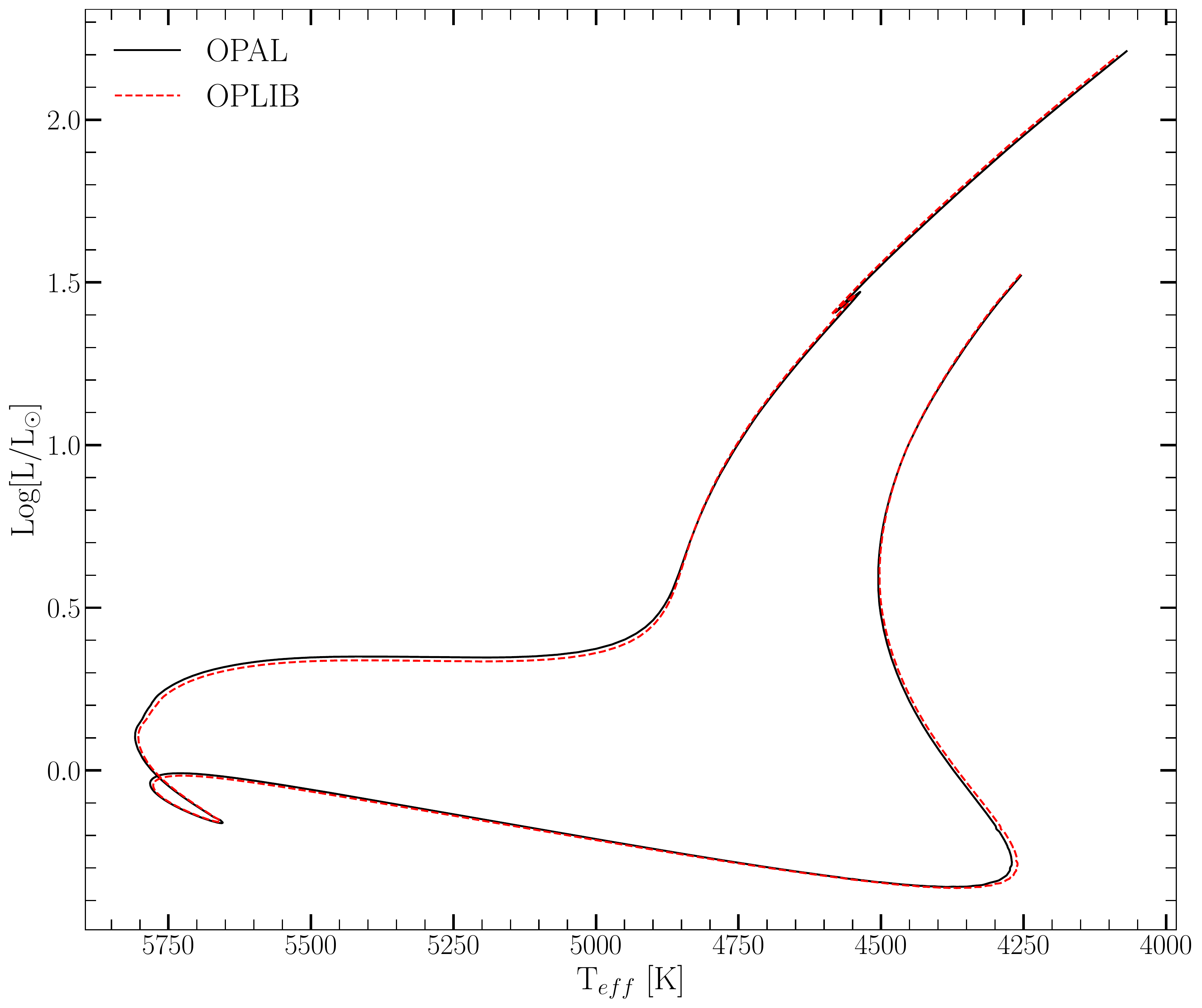}
	\caption{HR Diagram for the two SCSMs, OPAL and OPLIB. OPLIB is shown as a red
	dashed line.}
	\label{fig:OPLIBOPALHR}
\end{figure}

Solar calibrated stellar models evolved using GS98 OPAL and OPLIB opacity
tables (Figure \ref{fig:OPLIBOPALHR}) differ $\sim 0.5\%$ in the SCSM hydrogen
mass fractions and $\sim 1.5\%$ in the SCSM convective mixing length parameters
(Table \ref{tab:SCSMResults}). While the two evolutionary tracks are very
similar, note that the OPLIB SCSM's luminosity is systematically lower past the
solar age. While at the solar age the OPLIB SCSM luminosity is effectively the
same as the OPAL SCSM. This luminosity difference between OPAL and OPLIB based
models is not inconsistent with expectations given the more shallow radiative
temperature gradient resulting from the lower OPLIB opacities

\begin{table}
	\centering
	\begin{tabular}{l c c}
		\hline
		Model & $X$ & $\alpha_{ML}$ \\
		\hline
		\hline
		OPAL & 0.7066 & 1.9333 \\
		OPLIB & 0.7107 & 1.9629
	\end{tabular}
	\caption{Optimized parameters for SCSMs evolved using OPAL and OPLIB high
	temperature opacity tables.}
	\label{tab:SCSMResults}
\end{table}

\section{Modeling}\label{sec:modeling}
\begin{figure}
	\centering
	\includegraphics[width=0.45\textwidth]{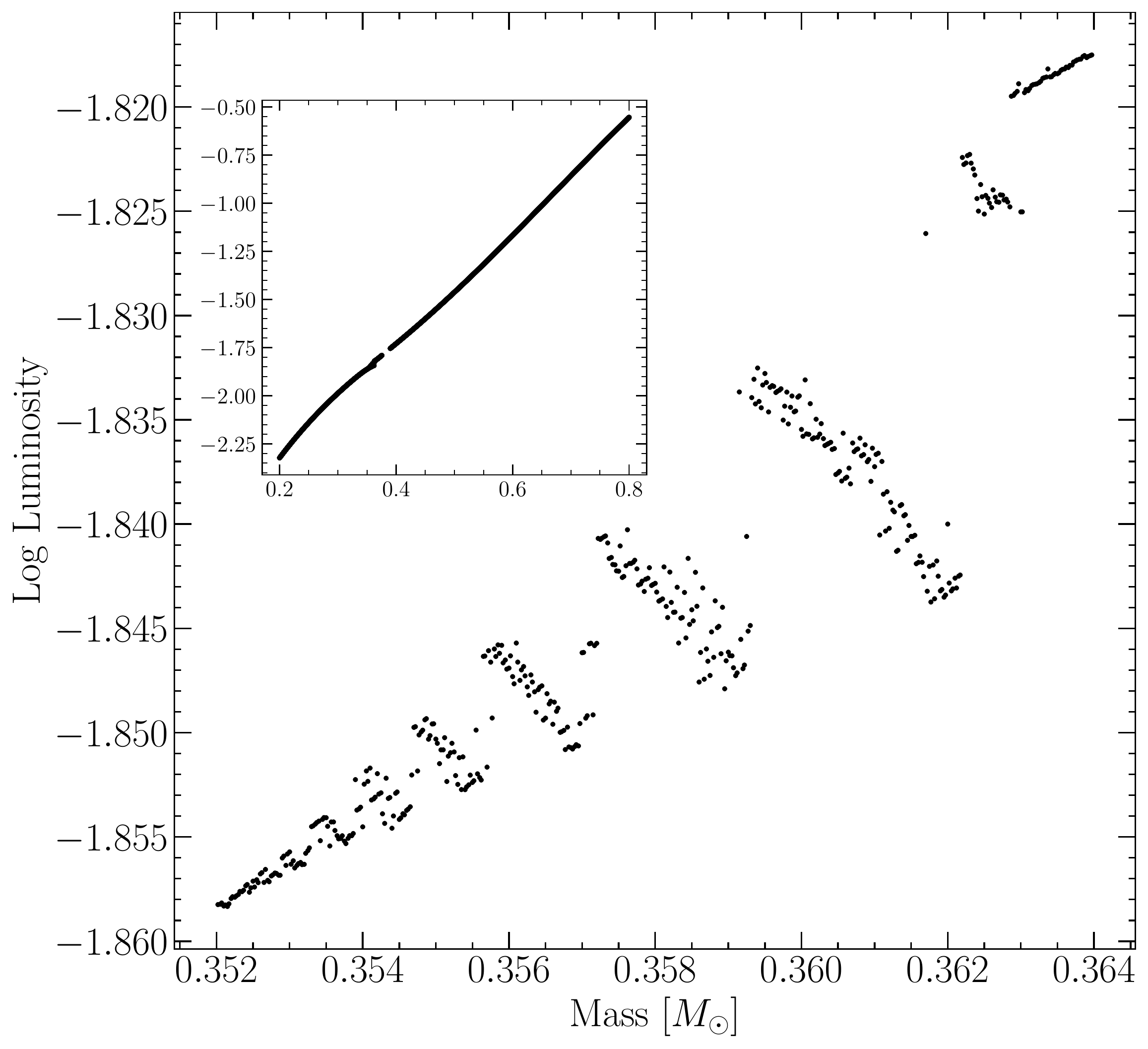}
	\includegraphics[width=0.45\textwidth]{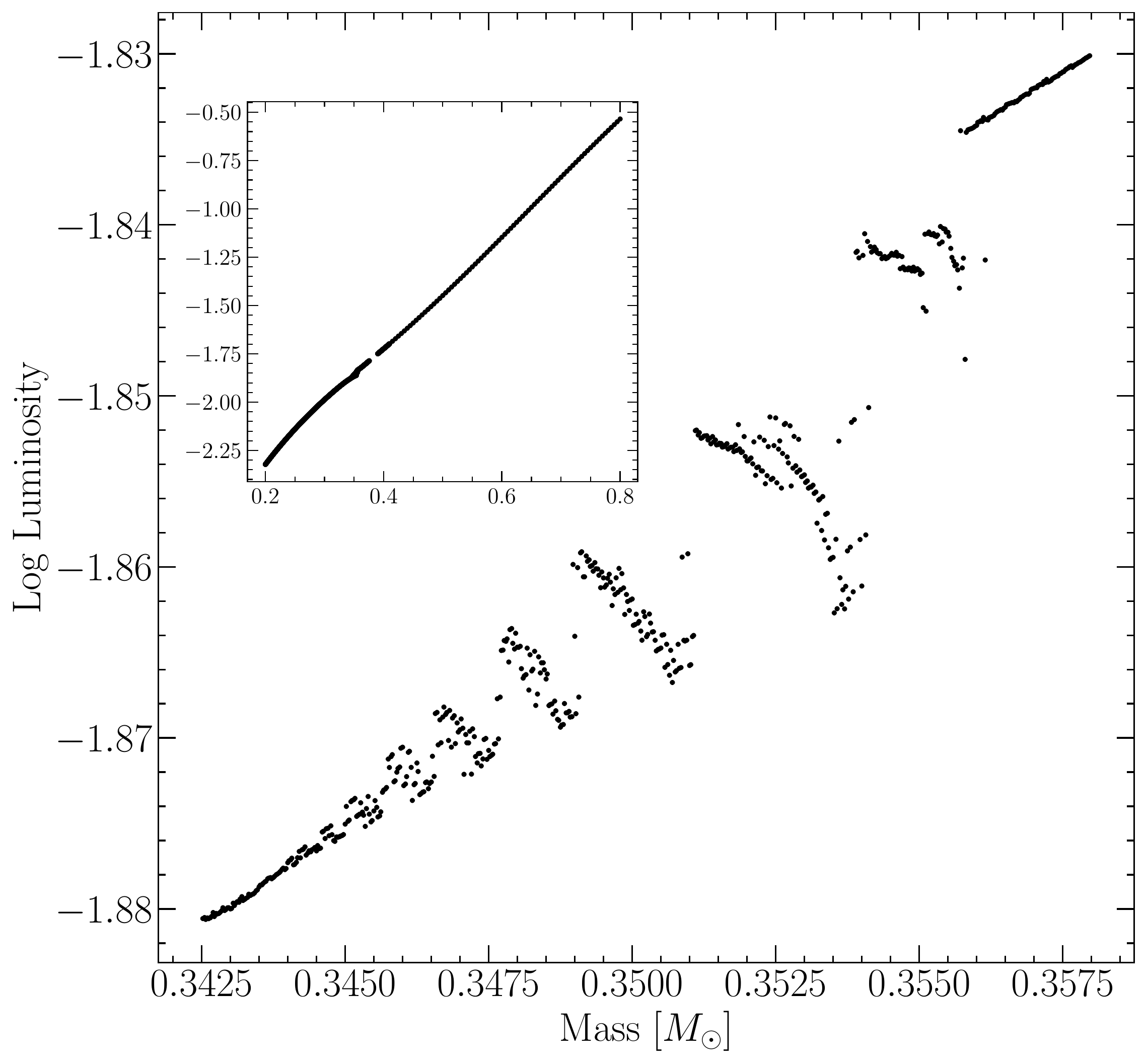}
	\caption{Mass-luminosity relation at 7 Gyrs for models evolved using OPAL opacity
	tables (top) and those evolved using OPLIB opacity tables (bottom). Note
	the lower mass range of the OPLIB Gap.}
	\label{fig:PunchIn}
		
\end{figure}

In order to model the Jao Gap we evolve two extremely finely sampled mass grids
of models. One of these grids uses the OPAL high-temperature opacity tables
while the other uses the OPLIB tables (Figure \ref{fig:PunchIn}). Each grid
evolves a model every 0.00025 $M_{\odot}$ from 0.2 to 0.4 $M_{\odot}$ and every
0.005 $M_{\odot}$ from 0.4 to 0.8 $M_{\odot}$. All models in both grids use a
GS98 solar composition, the (1, 101, 0) \texttt{FreeEOS} (version
{\color{red}2.7}) configuration, and 1000 year old pre-main sequence polytropic
models, with polytropic index 1.5, as their initial conditions. We
include gravitational settling in our models where elements are grouped
together. Finally, we set a maximum allowed timestep of 50 million years to
assure that we fully resolve the build of of core $^{3}$He in gap stars.

Despite the alternative view of convection provided by
\citet{MacDonald2018} discussed in Section \ref{sec:JaoGap}, given that the
mixing timescales in these low mass stars are so short \citep[between $10^{7}$s
and $10^{8}$s per][Figure 2 \& Equation 39, which present the
averaged velocity over the convection zone]{Jermyn2022} instantaneous mixing is a valid
approximation. Moreover, one principal motivation for a diffusive model of
convective mixing has been to account for a deuterium concentration gradient
which \citet{Chabrier1997} identify will develop when the deuterium lifetime
against proton capture is significantly shorter than the mixing timescale.
However, the treatment of energy generation used by DSEP \citep{Bahcall2001}
avoides this issue by computing both the equilibrium deuterium abundance and
luminosity of each shell individually, implicitly accounting for the overall
luminosity discrepancy identified by \citeauthor{Chabrier1997}.

Because in this work we are just interested in the location shift of the Gap as
the opacity source varies, we do not model variations in composition.
\citet{Mansfield2021,Jao2020,Feiden2021} all look at the effect composition has
on Jao Gap location. They find that as population metallicity increases so too
does the mass range and consequently the magnitude of the Gap. From an extremely
low metallicity population (Z=0.001) to a population with a more solar like
metallicity this shift in mass range can be up to 0.05 M$_{\odot}$
\citep{Mansfield2021}.

\subsection{Population Synthesis}
In order to compare the Gap to observations we use in house population
synthesis code. We empirically calibrate the relation between G, BP, and RP
magnitudes and their uncertainties along with the parallax/G magnitude
uncertainty relation using the Gaia Catalouge of Nearby Stas
\citep[GCNS,][]{GaiaCollaboration2021} and Equations \ref{eqn:plxCalib} \&
\ref{eqn:MagCalib}. $M_{g}$ is the Gaia G magnitude while $M_{i}$ is the
magnitude in the i$^\text{th}$ band, G, BP, or RP. The coefficients $a$, $b$,
and $c$ determined using a non-linear least squares fitting routine. Equation
\ref{eqn:plxCalib} then models the relation between G magnitude and parallax
uncertainty while Equation \ref{eqn:MagCalib} models the relation between each
magnitude and its uncertainty.

\begin{align}\label{eqn:plxCalib}
	\sigma_{plx}(M_{g}) = ae^{bM_{g}}+c
\end{align}
\begin{align}\label{eqn:MagCalib}
	\sigma_{i}(M_{i}) = ae^{M_{i}-b}+c
\end{align}

\noindent The full series of steps in our population synthesis code
are:

\begin{figure}
	\centering
	\includegraphics[width=0.45\textwidth]{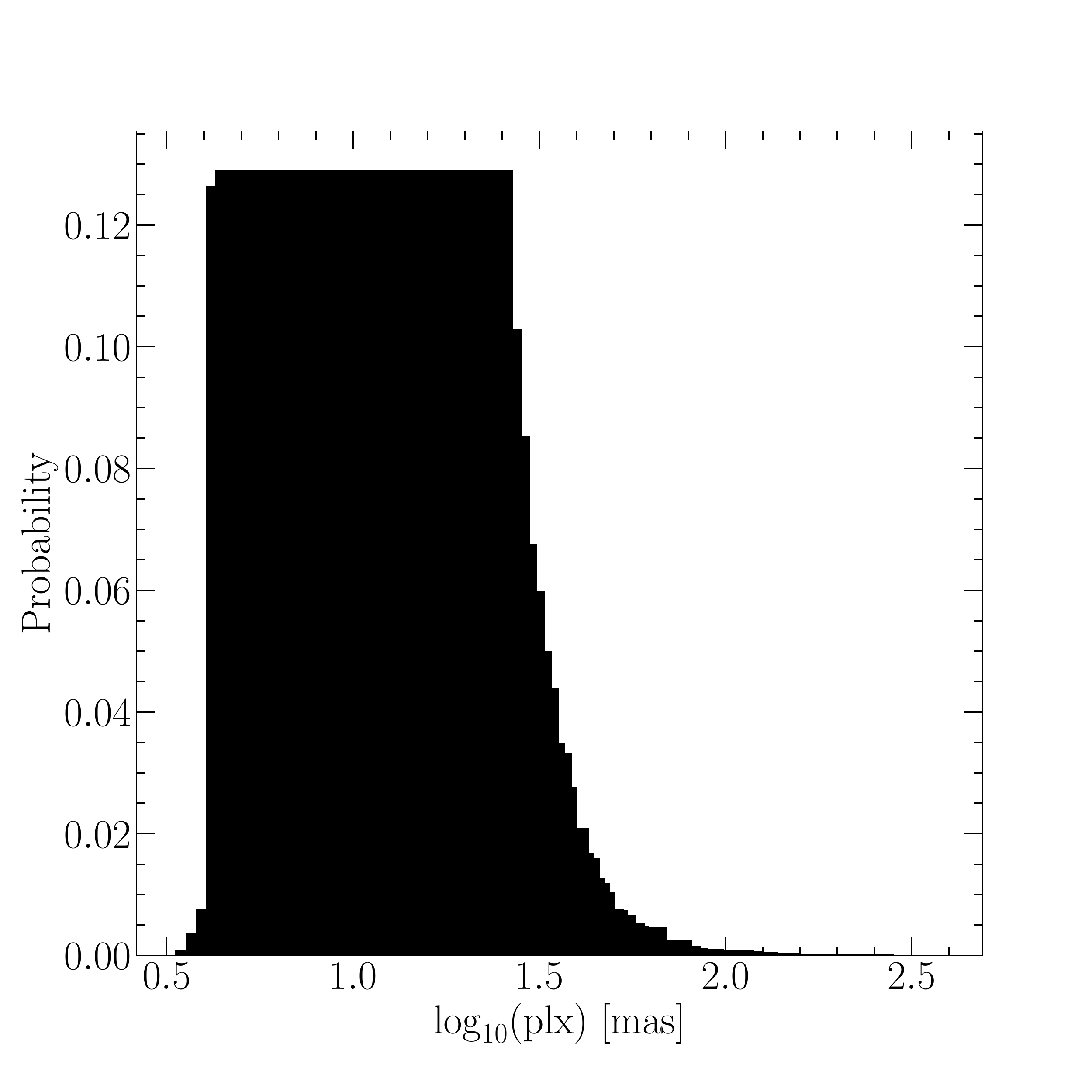}
	\caption{Probability distribution sampled when assigning true parallaxes to
	synthetic stars. This distribution is built from the GCNS and includes all
	stars with BP-RP colors between 2.3 and 2.9, the same color range
	of the Jao Gap.}
	\label{fig:pdist}
\end{figure}

\begin{enumerate}
	\item Sample from a \citet{Sollima2019} ($0.25 M_{\odot} < M < 1 M_{\odot}$,
		$\alpha=-1.34\pm0.07$) IMF to determine synthetic star mass.
	\item Find the closest model above and below the synthetic star, lineally
		interpolate these models' $T_{eff}$, $\log(g)$, and $\log(L)$ to those
		at the synthetic star mass.
	\item Convert synthetic star $g$, $T_{eff}$, and $Log(L)$ to Gaia G, BP,
		and RP magnitudes using the Gaia (E)DR3 bolometric corrections
		\citep{Creevey2022} along with code obtained thorough personal
		communication with Aaron Dotter \citep{Choi2016}.
	\item Sample from the GCNS parallax distribution (Figure \ref{fig:pdist}),
		limited to stars within the BP-RP color range of 2.3 -- 2.9, to assign
		synthetic star a ``true'' parallax.
	\item Use the true parallax to find an apparent magnitude for each filter.
	\item Evaluate the empirical calibration given in Equation
		\ref{eqn:plxCalib} to find an associated parallax uncertainty. Then
		sample from a normal distribution with a standard deviation equal to
		that uncertainty to adjust the true parallax resulting in an
		``observed'' parallax.
	\item Use the ``observed'' parallax and the apparent magnitude to find an
		``observed'' magnitude.
	\item Fit the empirical calibration given in Equation \ref{eqn:MagCalib} to
		the GCNS and evaluate it to give a magnitude uncertainty scale in each
		band.
	\item Adjust each magnitude by an amount sampled from a normal
		distribution with a standard deviation of the magnitude uncertainty
		scale found in the previous step.
\end{enumerate}

\begin{figure*}
	\centering
	\includegraphics[width=0.85\textwidth]{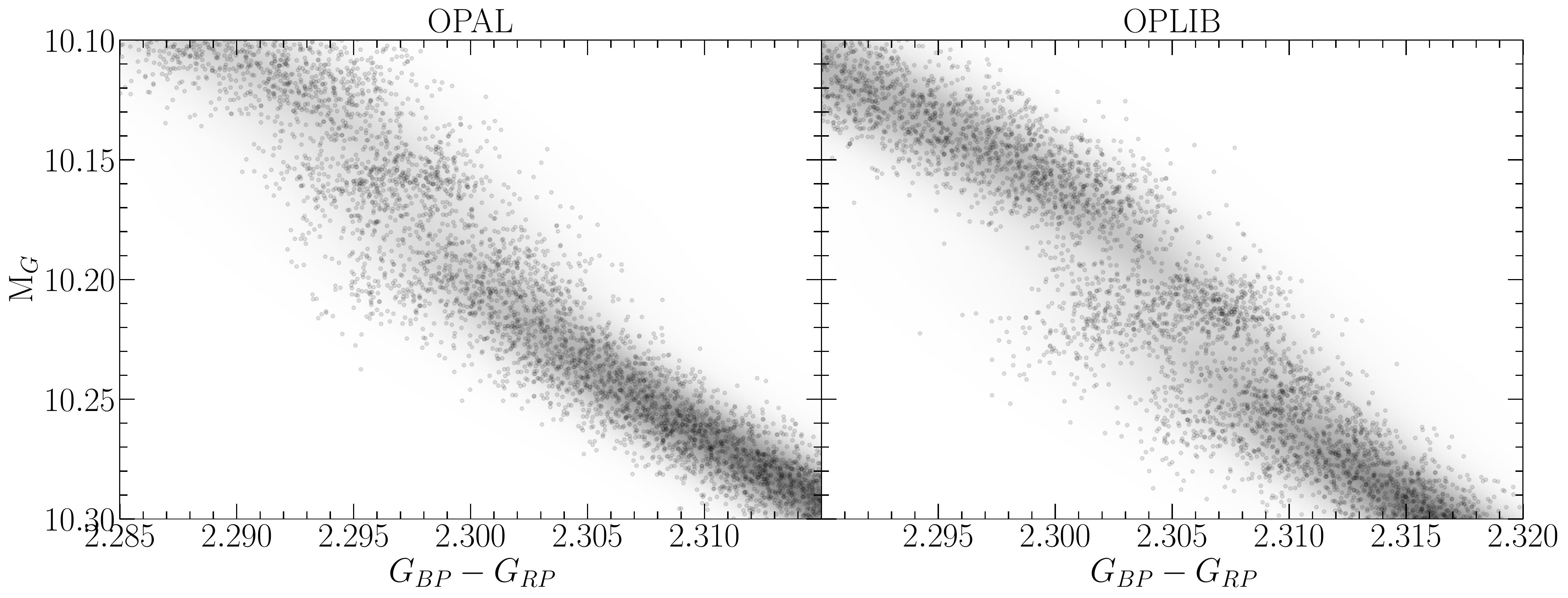}
	\caption{Population synthesis results for models evolved with OPAL (left)
	and models evolved with OPLIB (right). A Gaussian kernel-density estimate
	has been overlaid to better highlight the density variations.}
	\label{fig:PopSynthCompareBasic}
\end{figure*}

This method then incorporates both photometric and astrometric uncertainties
into our population synthesis. An example 7 Gyr old synthetic populations
using OPAL and OPLIB opacities are presented in Figure
\ref{fig:PopSynthCompareBasic}.

\subsection{Mixing Length Dependence}
In order to test the sensitivity of Gap properties to mixing length we
evolve three separate sets OPLIB of models. The first uses a GS98
solar calibrated mixing length, the second uses a mixing length of
1.5, and the third uses a mixing length of 1.0.

\begin{figure}
	\centering
	\includegraphics[width=0.45\textwidth]{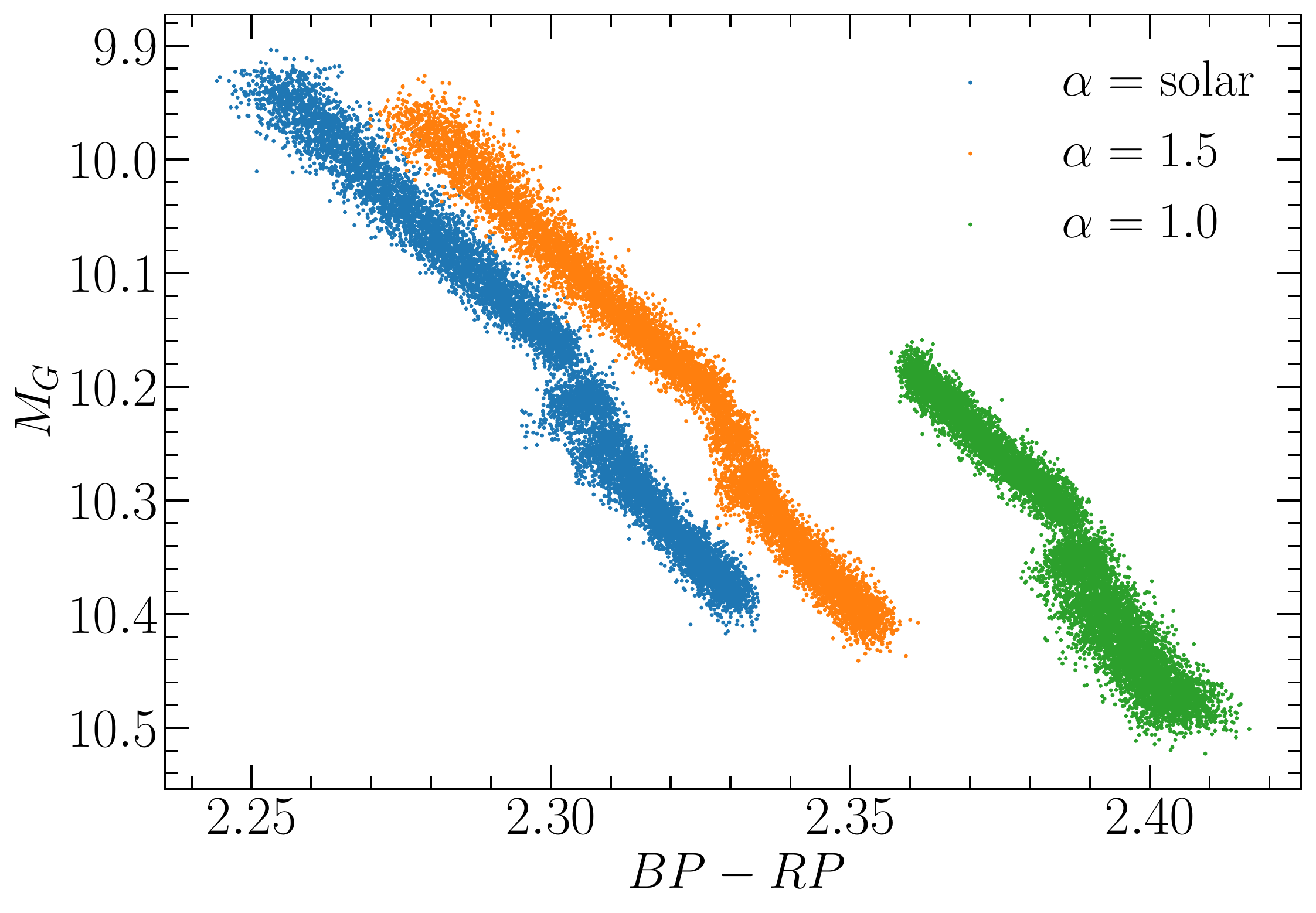}
	\caption{CMD showing OPLIB populations (from left to right) A, B, and C.}
	\label{fig:MixingLengthCMD}
\end{figure}

\begin{figure}
	\centering
	\includegraphics[width=0.45\textwidth]{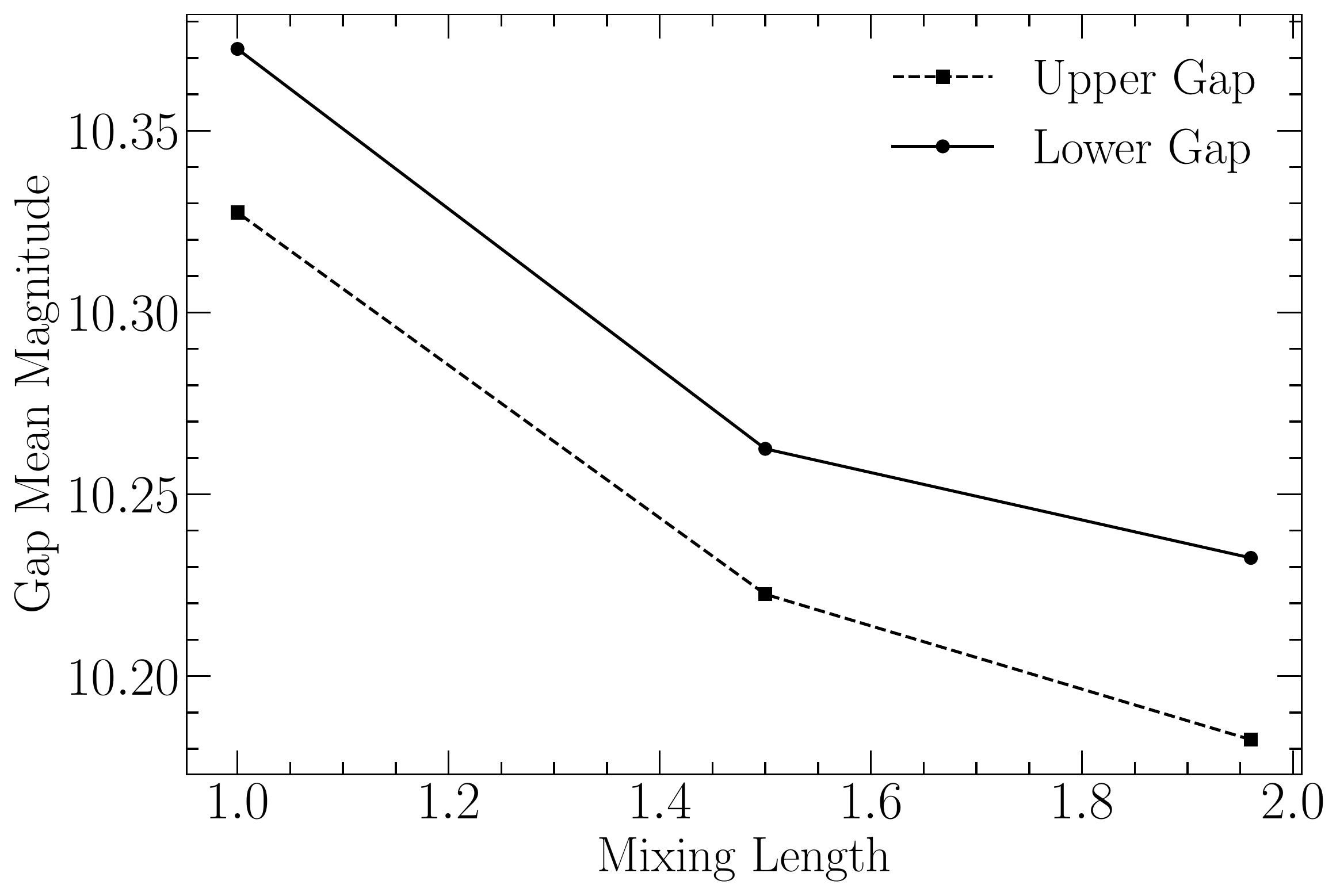}
	\caption{Location of the two identified paucities of stars in OPLIB synthetic
	populations as a function of the mixing length used.}
	\label{fig:MixingLengthScaling}
\end{figure}

We find a clear inverse correlation between mixing length parameter used and
the magnitude of the Jao Gap Figures \ref{fig:MixingLengthCMD} \&
\ref{fig:MixingLengthScaling} ($\mu_{G} \propto -1.5\alpha_{ML}$, where
$\mu_{G}$ is the mean magnitude of the Gap). This is somewhat surprising given
the long established view that the mixing length parameter is of little
relevance in fully convective stars \citep{Baraffe1997}. We find an approximate
0.3 magnitude shift in both the color and magnitude comparing a solar
calibrated mixing length to a mixing length of 1.5, despite only a 16K
difference in effective temperature at 7Gyr between two 0.3 solar mass models.
The slight temperature differences between these models are
attributable to the steeper adiabatic temperature gradients just below the
atmosphere in the solar calibrated mixing length model compared to the
$\alpha_{ML} = 1.5$ model ($\nabla_{ad,solar} - \nabla_{ad,1.5} \approx 0.05$).
Despite this relatively small temperature variance, the large magnitude
difference is expected due to the extreme sensitivity of the bolometric
corrections on effective temperature at these low temperatures. The
mixing length then provides a free parameter which may be used to shift the gap
location in order to better match observations without having a major impact on
the effective temperature of models. Moreover, recent work indicates that using
a solar calibrated mixing length is not appropriate for all stars
\citep[e.g.][]{Trampedach2014, Joyce2018}.

Given the variability of gap location with mixing length, it is possible that a
better fit to the gap location may be achieved through adjustment of the
convective mixing length parameter. However, calibrations of the mixing length
for stars other than the sun have focused on stars with effective temperature
at or above that of the sun and there are no current calibrations of the mixing
length parameter for M dwarfs. Moreover, there are additional uncertainties
when comparing the predicted gap location to the measured gap location, such as
those in the conversion from effective temperature, surface gravity, and
luminosity to color, which must be considered if the mixing length is to be
used as a gap location free parameter. Given the dangers of freely adjustable
parameters and the lack of an a priori expectation for what the convective
mixing parameter should be for the population of M Dwarfs in the Gaia DR2 and
EDR3 CMD any attempt to use the Jao Gap magnitude to calibrate a mixing length
value must be done with caution, and take into account the other uncertainties
in the stellar models which could affect the Jao Gap magnitude.

\section{Results}\label{sec:results}
\begin{figure*}[ht!]
	\centering
	\includegraphics[width=0.45\textwidth]{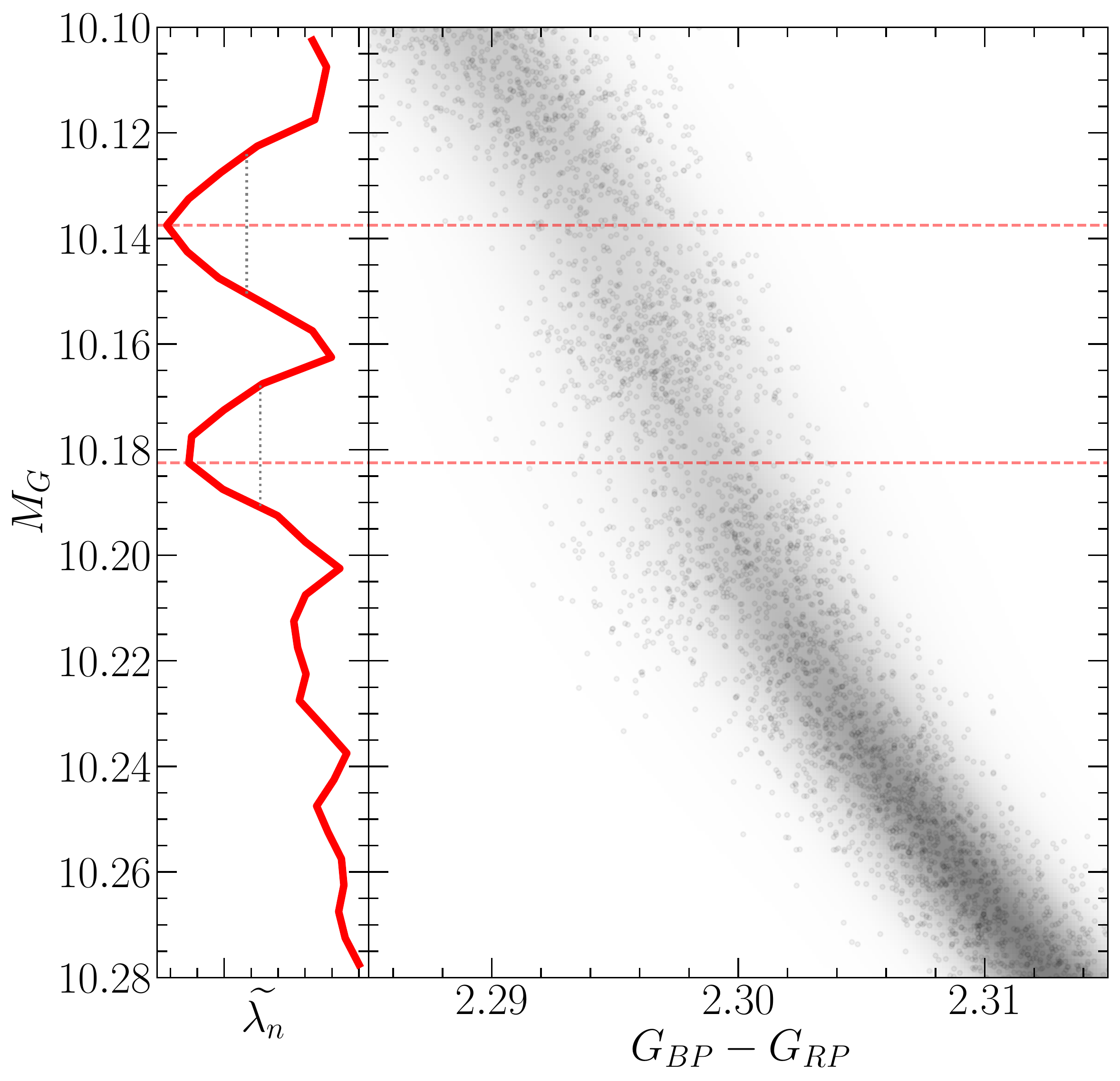}
	\includegraphics[width=0.45\textwidth]{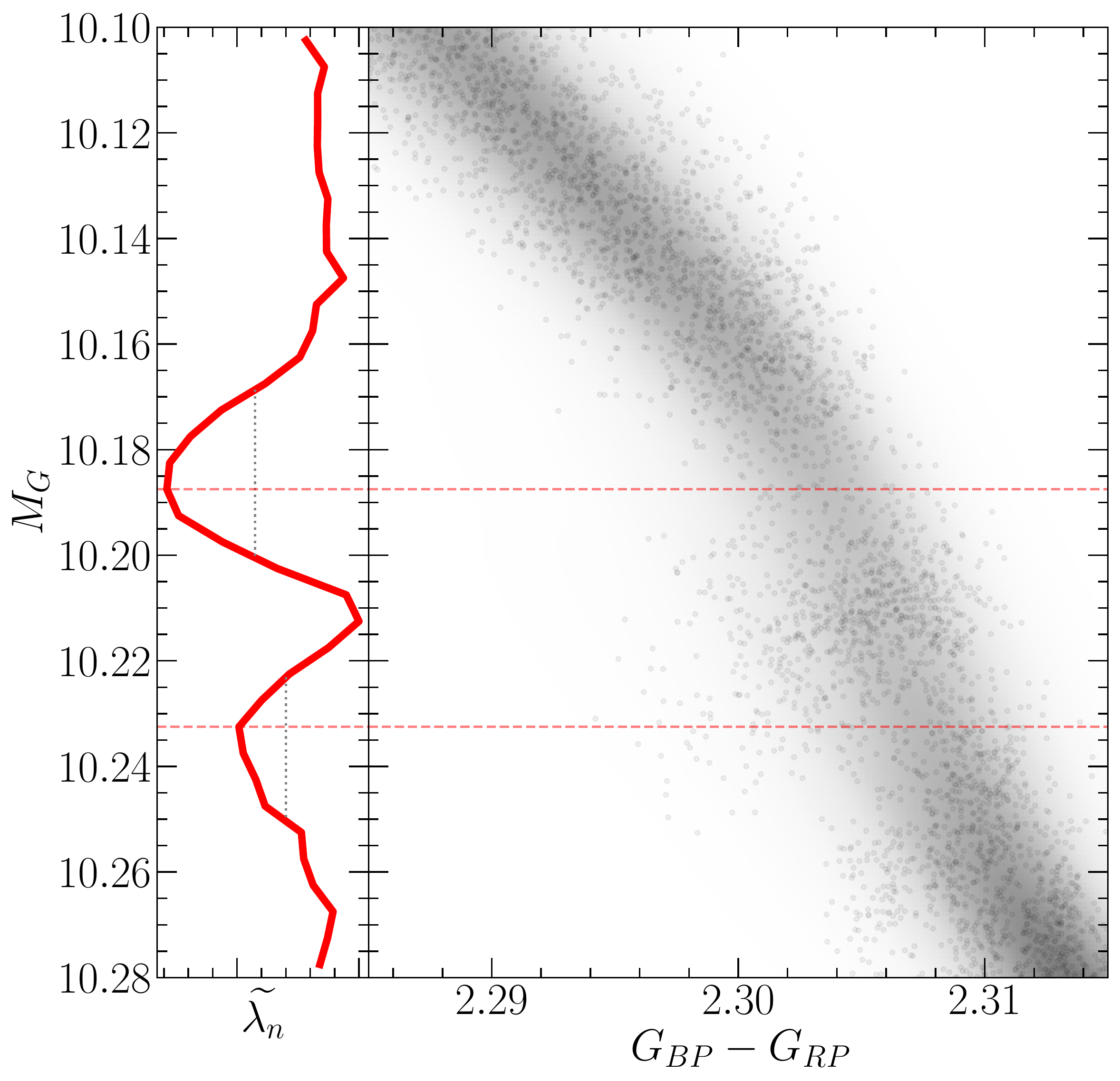}
	\caption{(right panels) OPAL (left) and OPLIB (right) synthetic
	populations. (left panels) Normalized linear number density along the
	magnitude axis. A dashed line has been extended from the peak through both
	panels to make clear where the identified Jao Gap location is wrt. to the
	population. }
	\label{fig:JaoGapLocator}
\end{figure*}

We quantify the Jao Gap location along the magnitude (Table
\ref{tab:GapLocation}) axis by sub-sampling our synthetic populations, finding
the linear number density along the magnitude axis of each sub-sample,
averaging these linear number densities, and extracting any peaks above a
prominence threshold of 0.1 as potential magnitudes of the Jao Gap (Figure
\ref{fig:JaoGapLocator}). Gap widths are measured at 50\% the height of the peak
prominence. We use the python package \texttt{scipy} \citep{2020SciPy-NMeth} to
both identify peaks and measure their widths. 

\begin{table}
	\centering
	\begin{tabular}{c | c c c}
		\hline
		Model & Location & Prominence & Width\\
		\hline
		\hline
		OPAL 1 & 10.138 & 0.593 & 0.027 \\
		OPAL 2 & 10.183 & 0.529 & 0.023 \\
		OPLIB 1 & 10.188 & 0.724 & 0.032 \\
		OPLIB 2 & 10.233 & 0.386 & 0.027 
	\end{tabular}
	\caption{Locations identified as potential Gaps.}
	\label{tab:GapLocation}
\end{table}

In both OPAL and OPLIB synthetic populations our Gap identification method
finds two gaps above the prominence threshold. The identification of more than
one gap is not inconsistent with the mass-luminosity relation seen in the grids
we evolve. As noise is injected into a synthetic population smaller features will
be smeared out while larger ones will tend to persist. The mass-luminosity
relations shown in in Figure \ref{fig:PunchIn} make it clear that there are: (1),
multiple gaps due to stars of different masses undergoing convective mixing
events at different ages, and (2), the gaps decrease in width moving to lower
masses / redder. Therefore, the multiple gaps we identify are attributable to
the two bluest gaps being wide enough to not smear out with noise. In fact, if
we lower the prominence threshold just slightly from 0.1 to 0.09 we detect a
third gap in both the OPAL and OPLIB datasets where one would be expected.

Previous modeling efforts \citep[e.g.][]{Feiden2021} have not identified
multiple gaps. This is likely due to two reasons: (1), previous studies have
allowed metallicity to vary across their model grids, further smearing the gaps
out, and (2), previous studies have used more coarse underlying mass grids,
obscuring features smaller than their mass step. While this dual-gap structure
has not been seen in models before, a more complex gap structure is not totally
unprecedented as \citet{Jao2021} identifies an additional under-dense region
below the primary gap in EDR3 data. As part of a follow up series of papers, we
are conducting further work to incorporate metallicity variations while still
using the finer mass sampling presented here.

The mean gap location of the OPLIB population is at a fainter magnitude than
the mean gap location of the OPAL population. Consequently, in the OPLIB sample
the convective mixing events which drive the kissing instability begin
happening at lower masses (i.e. the convective transition mass decreases). A
lower mass range will naturally result in a fainter mean gap magnitude.

\begin{figure}
	\centering
	\includegraphics[width=0.45\textwidth]{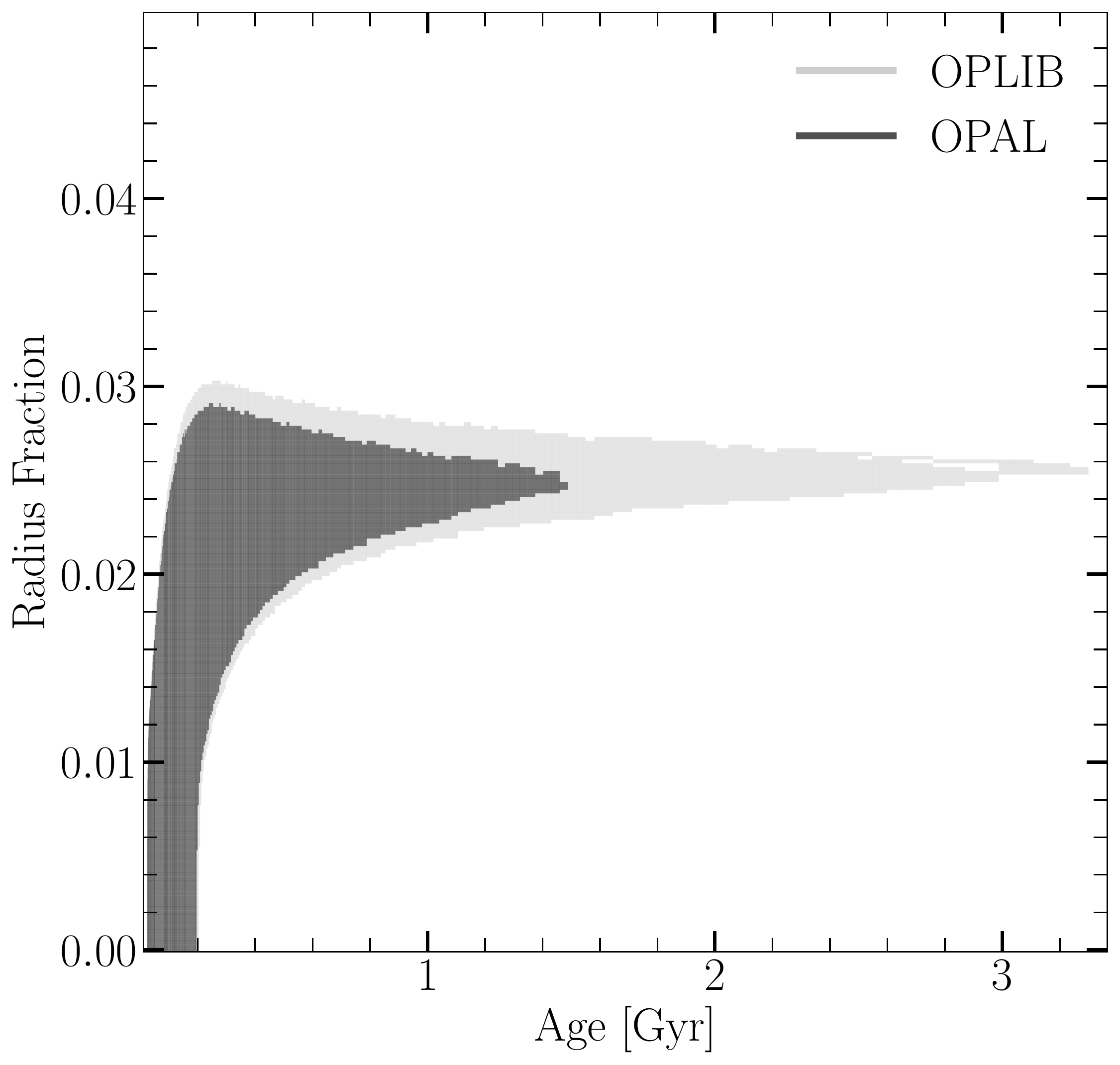}
	\caption{Portions of 0.3526 $M_{\odot}$ OPAL and OPLIB stellar models
	showing the interior shells which are radiative (black region). Note that
	for clarity only one convective mixing event from each model is shown. Note
	how the radiative zone in the OPLIB model is larger.}
	\label{fig:Unstable}
\end{figure}

\begin{figure}
	\centering
	\includegraphics[width=0.45\textwidth]{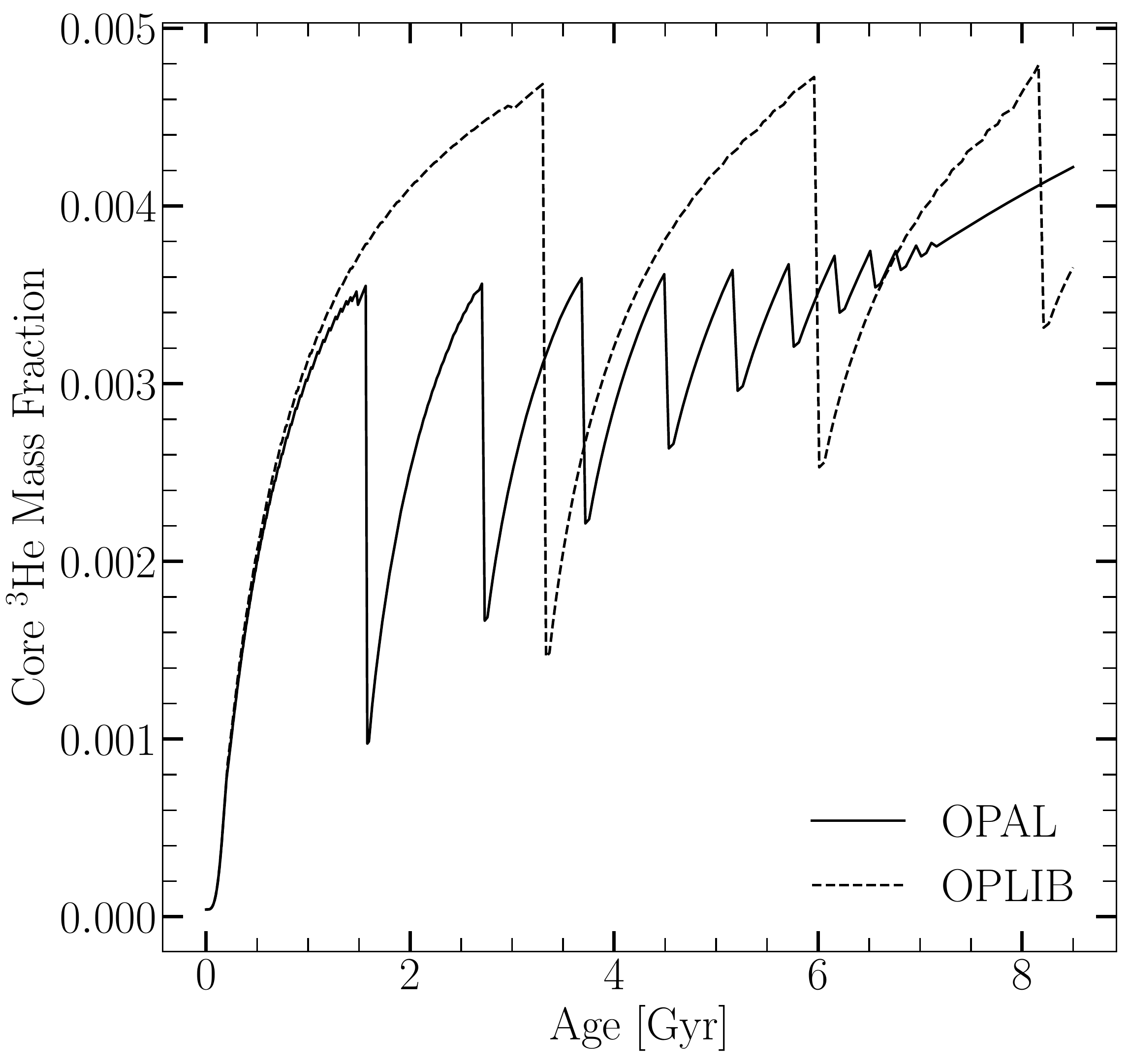}
	\caption{Core $^{3}$He mass fraction for  0.3526 $M_{\odot}$ models evolved
	with OPAL and OPLIB (within the Jao Gap's mass range for both). Note how
	the OPLIB model's core $^{3}$He mass fraction grows at approximately the
	same rate as the OPAL model's but continues uninterrupted for longer.}
	\label{fig:OPALOPLIB3He}
\end{figure}

Mixing events at lower masses in OPLIB models are attributable to the radially
thicker, at the same mass, radiative zones (Figure \ref{fig:Unstable}). This
thicker radiative zone will take more time to break down and is characteristic
of OPLIB models as of a result of their slightly lower opacities. A lower
opacity fluid will have a more shallow radiative temperature gradient than a
higher opacity fluid; however, as the adiabatic temperature gradient remains
essentially unchanged as a function of radius, a larger interior radius of the
model will remain unstable to radiation. This thicker radiative zone will
increase the time it takes the core convective zone to meet up with convective
envelope meaning that lower mass models can sustain a radiative zone for
longer than they could otherwise; thus; lower opacities push the convective
transition mass down. We can additionally see this longer lived radiative zone
in the core $^{3}$He mass fraction, in which OPLIB models reach much higher
concentrations --- at approximately the same growth rate --- for the same mass
as OPAL models do (Figure \ref{fig:OPALOPLIB3He}). 

The most precise published Gap location comes from \citet{Jao2020} who use EDR3
to locate the Gap at $M_{G} \sim 10.3$, we identify the Gap at a similar
location in the GCNS data. The Gap in populations evolved using OPLIB
tables is closer to this measurement than it is in populations evolved using
OPAL tables (Table \ref{tab:GapLocation}). It should be noted that the exact
location of the observed Gap is poorly captured by a single value as the Gap
visibly compresses across the width of the main-sequence, wider on the blue
edge and narrower on the red edge such that the observed Gap has downward
facing a wedge shape (Figure \ref{fig:JaoGap}). This wedge shape is not
successfully reproduced by either any current models or the modeling we preform
here. We elect then to specify the Gap location where this wedge is at its
narrowest, on the red edge of the main sequence.

The Gaps identified in our modeling have widths of approximately 0.03
magnitudes, while the shift from OPAL to OPLIB opacities is 0.05 magnitudes.
With the prior that the Gaps clearly shift before noise is injected we know
that this shift is real. However, the shift magnitude and Gap width are of
approximately the same size in our synthetic populations. Moreover,
\citet{Feiden2021} identify that the shift in the modeled Gap mass from [Fe/H]
= 0 to [Fe/H] = +0.5 as 0.04$M_{\odot}$, whereas we only see an approximate
$0.01$ M$_{\odot}$ shift between OPAL and OPLIB models. Therefore, the
Gap location will likely not provide a usable constraint on the opacity
source.

\section{Conclusion}\label{sec:conclusion}
%
%
The Jao Gap provides an intriguing probe into the interior physics of M Dwarfs
stars where traditional methods of studying interiors break down. However,
before detailed physics may be inferred it is essential to have models which
are well matched to observations. Here we investigate whether the OPLIB opacity
tables reproduce the Jao Gap location and structure more accurately than the
widely used OPAL opacity tables. We find that while the OPLIB tables do shift
the Jao Gap location more in line with observations, by approximately 0.05
magnitudes, the shift is small enough that it is likely not distinguishable
from noise due to population age and chemical variation. However, future
measurement of [Fe/H] for stars within the gap will be helpful in constraining
the degree to which the gap should be smeared by these theoretical models.

We also find that both the color and magnitude of the Jao Gap are
correlated to the convective mixing length parameter. Specifically, a lower
mixing length parameter will bring the gap in the populations presented in this
paper more in line with the current best estimate for the actual gap magnitude.
Using this relation it may be possible for mixing length to be calibrated for
low mass stars such that models match the Jao Gap location. Further, the Jao
gap location may provide a test of alternative convection models such as
entropy calibrated convection \citep{Spada2021}. Both of these potential uses
require careful handeling of other uncertanties such as the uncertanties in
bolometric correction, popupulation composition, and population age. As we
currently do not have reason to suspect that the mixing length for the low mass
stars in the DR2 and ERD3 CMD is substantially lower than that of the sun we
leave the investigation of these potential additionl uses for future work.

Finally, we do not find that the OPLIB opacity tables help in reproducing the
as yet unexplained wedge shape of the observed Gap.

\acknowledgments{
	This work has made use of the NASA astrophysical data system (ADS). We
	would like to thank Elisabeth Newton, Aaron Dotter, and Gregory Feiden for
	their support and for useful discussion related to the topic of this paper.
	We would like to thank our reviewer for their critical eye and their
	guidance to investigate to effects of the mixing length on the Gap
	Location. Additionally, we would like to thank James Colgan and the Los
	Alamos T-1 group for their assistance with the OPLIB opacity tables and
	support for the public release of \texttt{pyTOPSScrape}. We acknowledge the
	support of a NASA grant (No. 80NSSC18K0634). 
}
\acknowledgments


\software{
	The Dartmouth Stellar Evolution Program (DSEP) \citep{Dotter2008},
	\texttt{BeautifulSoup} \citep{richardson2007beautiful},
	\texttt{mechanize} \citep{chandra2015python},
	\texttt{FreeEOS} \citep{Irwin2012},
	\texttt{pyTOPSScrape} \citep{Boudreaux22}
}

\appendix

\section{\texttt{pyTOPSScrape}}\label{apx:pytopsscrape}
\texttt{pyTOPSScrape} provides an easy to use command line and python interface
for the OPLIB opacity tables accessed through the TOPS web form. Extensive
documentation of both the command line and programmatic interfaces is linked
in the version controlled repository. However, here we provide a brief,
illustrative, example of potential use.

Assuming \texttt{pyTOPSScrape} has been installed and given some working
directory which contains a file describing a base composition (``comp.dat'')
and another file containing a list of rescalings of that base composition
(``rescalings.dat'') (both of these file formats are described in detail in the
documentation), one can query OPLIB opacity tables and convert them to a form
mimicking that of type 1 OPAL high temperature opacity tables using the
following shell command.

\begin{verbatim}
	$ generateTOPStables comp.dat rescalings.dat -d ./TOPSCache -o out.opac -j 20
\end{verbatim}

\noindent For further examples of pyTOPSScrape please visit the repository.

\section{Interpolating $\rho \rightarrow $ R}\label{apx:interp}
OPLIB parameterizes $\kappa_{R}$ as a function of mass density, temperature in keV,
and composition. Type 1 OPAL high temperature opacity tables, which DSEP and
many other stellar evolution programs use, instead parameterizes opacity as a function
of temperature in Kelvin, $R$ (Equation \ref{eqn:Req}), and composition. The
conversion from temperature in keV to Kelvin is trivial (Equation
\ref{eqn:K2Kev}).
\begin{align}\label{eqn:Req}
	R = \frac{\rho}{T_{6}^{3}}
\end{align}
\begin{align}\label{eqn:K2Kev}
	T_{K} = T_{keV} * 11604525.0061657
\end{align}
However, the conversion from mass density to $R$ is more involved. Because $R$
is coupled with both mass density and temperature there there is no way to
directly convert tabulated values of opacity reported in the OPLIB tables to
their equivalents in $R$ space. The TOPS webform does allow for a
density range to be specified at a specific temperature, which allows for R
values to be directly specified. However, issuing a query to the TOPS webform
for not just every composition in a Type 1 OPAL high temperature opacity table
but also every temperature for every composition will increase the number of
calls to the webform by a factor of 70. Therefore, instead of directly
specifying R through the density range we choose to query tables over a
broad temperature and density range and then rotate these tables,
interpolating $\kappa_{R}(\rho,T_{eff}) \rightarrow \kappa_{R}(R,T_{eff})$.

To preform this rotation we use the \texttt{interp2d} function within
\texttt{scipy}'s \texttt{interpolate} \citep{2020SciPy-NMeth} module to
construct a cubic bivariate B-spline \citep{Dierckx1981} interpolating function
$s$, with a smoothing factor of 0, representing the surface $\kappa_{R}(\rho,
T_{eff})$. For each $R^{i}$ and $T^{j}_{eff}$ reported in type 1 OPAL tables,
we evaluate Equation \ref{eqn:Req} to find $\rho^{ij} =
\rho(T^{j}_{eff},R^{i})$.  Opacities in $T_{eff}$, $R$ space are then inferred
as $\kappa^{ij}_{R}(R^{i},T^{j}_{eff}) = s(\rho^{ij}, T^{j}_{eff})$. 

\begin{figure}
	\centering
	\includegraphics[width=0.45\textwidth]{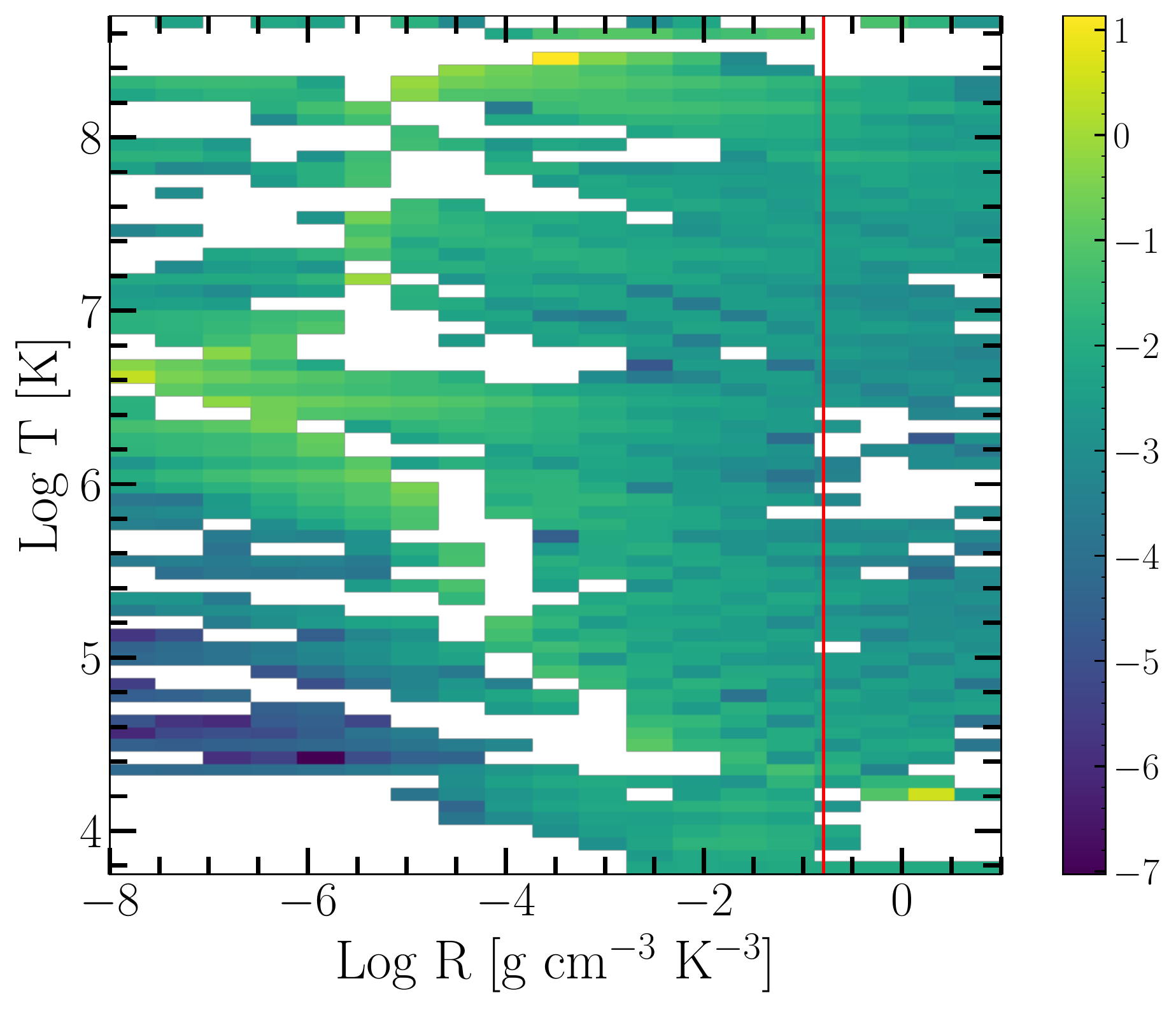}
	\caption{Log Fractional Difference between opacities in $\kappa_{R}(\rho,
	T_{eff})$ space directly queried from the OPLIB web-form and those which
	have been interpolated into $\log(R)$ space and back. Note that, due to the
	temperature grid of type 1 OPAL tables not aligning perfectly which the temperature
	grid OPLIB uses there may be edge effects where the interpolation is poorly
	constrained. The red line corresponds to $\log(R) = -0.79$ where much of a
	stellar model's radius exists.}
	\label{fig:fracdiff}
\end{figure}

As first-order validation of this interpolation scheme we can preform a similar
interpolation in the opposite direction, rotating the tables back to
$\kappa_{R}(\rho, T_{eff})$ and then comparing the initial, ``raw'', opacities
to those which have gone through the interpolations process. Figure
\ref{fig:fracdiff} shows the fractional difference between the raw opacities
and a set which have gone through this double interpolation. The red line
denotes $\log(R)=-0.79$ where models near the Jao Gap mass range will
tend to sit for much of their radius. Along the $\log(R)=-0.79$ line the mean
fractional difference is $\langle \delta \rangle = 0.005$ with an uncertainty of
$\sigma_{\langle\delta\rangle} = 0.013$. One point of note is that, because the
initial rotation into $\log(R)$ space also reduces the domain of the opacity
function, interpolation-edge effects which we avoid initially by extending the
domain past what type 1 OPAL tables include cannot be avoided when
interpolating back into $\rho$ space. 

\bibliography{ms}{}

\begin{thebibliography}{}
\expandafter\ifx\csname natexlab\endcsname\relax\def\natexlab#1{#1}\fi
\providecommand{\url}[1]{\href{#1}{#1}}
\providecommand{\dodoi}[1]{doi:~\href{http://doi.org/#1}{\nolinkurl{#1}}}
\providecommand{\doeprint}[1]{\href{http://ascl.net/#1}{\nolinkurl{http://ascl.net/#1}}}
\providecommand{\doarXiv}[1]{\href{https://arxiv.org/abs/#1}{\nolinkurl{https://arxiv.org/abs/#1}}}

\bibitem[{{Amard} {et~al.}(2019){Amard}, {Palacios}, {Charbonnel}, {Gallet},
  {Georgy}, {Lagarde}, \& {Siess}}]{Amard2019}
{Amard}, L., {Palacios}, A., {Charbonnel}, C., {et~al.} 2019, \aap, 631, A77,
  \dodoi{10.1051/0004-6361/201935160}

\bibitem[{{Bahcall} {et~al.}(2001){Bahcall}, {Pinsonneault}, \&
  {Basu}}]{Bahcall2001}
{Bahcall}, J.~N., {Pinsonneault}, M.~H., \& {Basu}, S. 2001, \apj, 555, 990,
  \dodoi{10.1086/321493}

\bibitem[{{Bahcall} {et~al.}(2005){Bahcall}, {Serenelli}, \&
  {Basu}}]{Bahcall2005}
{Bahcall}, J.~N., {Serenelli}, A.~M., \& {Basu}, S. 2005, \apjl, 621, L85,
  \dodoi{10.1086/428929}

\bibitem[{{Baraffe} \& {Chabrier}(2018)}]{Baraffe2018}
{Baraffe}, I., \& {Chabrier}, G. 2018, \aap, 619, A177,
  \dodoi{10.1051/0004-6361/201834062}

\bibitem[{{Baraffe} {et~al.}(1997){Baraffe}, {Chabrier}, {Allard}, \&
  {Hauschildt}}]{Baraffe1997}
{Baraffe}, I., {Chabrier}, G., {Allard}, F., \& {Hauschildt}, P.~H. 1997, \aap,
  327, 1054.
\newblock \doarXiv{astro-ph/9704144}

\bibitem[{Boudreaux(2022)}]{Boudreaux22}
Boudreaux, E. 2022, tboudreaux/pytopsscrape: pyTOPSScrape v1.0, v1.0,  Zenodo,
  \dodoi{10.5281/zenodo.7094198}

\bibitem[{{Chabrier} \& {Baraffe}(1997)}]{Chabrier1997}
{Chabrier}, G., \& {Baraffe}, I. 1997, \aap, 327, 1039.
\newblock \doarXiv{astro-ph/9704118}

\bibitem[{Chandra \& Varanasi(2015)}]{chandra2015python}
Chandra, R.~V., \& Varanasi, B.~S. 2015, Python requests essentials (Packt
  Publishing Ltd)

\bibitem[{{Choi} {et~al.}(2016){Choi}, {Dotter}, {Conroy}, {Cantiello},
  {Paxton}, \& {Johnson}}]{Choi2016}
{Choi}, J., {Dotter}, A., {Conroy}, C., {et~al.} 2016, \apj, 823, 102,
  \dodoi{10.3847/0004-637X/823/2/102}

\bibitem[{{Colgan} {et~al.}(2016){Colgan}, {Kilcrease}, {Magee}, {Sherrill},
  {Abdallah}, {Hakel}, {Fontes}, {Guzik}, \& {Mussack}}]{Colgan2016}
{Colgan}, J., {Kilcrease}, D.~P., {Magee}, N.~H., {et~al.} 2016, in APS Meeting
  Abstracts, Vol. 2016, APS Division of Atomic, Molecular and Optical Physics
  Meeting Abstracts, D1.008

\bibitem[{{Creevey} {et~al.}(2022){Creevey}, {Sordo}, {Pailler}, {Fr{\'e}mat},
  {Heiter}, {Th{\'e}venin}, {Andrae}, {Fouesneau}, {Lobel}, {Bailer-Jones},
  {Garabato}, {Bellas-Velidis}, {Brugaletta}, {Lorca}, {Ordenovic}, {Palicio},
  {Sarro}, {Delchambre}, {Drimmel}, {Rybizki}, {Torralba Elipe}, {Korn},
  {Recio-Blanco}, {Schultheis}, {De Angeli}, {Montegriffo}, {Abreu Aramburu},
  {Accart}, {{\'A}lvarez}, {Bakker}, {Brouillet}, {Burlacu}, {Carballo},
  {Casamiquela}, {Chiavassa}, {Contursi}, {Cooper}, {Dafonte}, {Dapergolas},
  {de Laverny}, {Dharmawardena}, {Edvardsson}, {Le Fustec},
  {Garc{\'\i}a-Lario}, {Garc{\'\i}a-Torres}, {Gomez},
  {Gonz{\'a}lez-Santamar{\'\i}a}, {Hatzidimitriou}, {Jean-Antoine Piccolo},
  {Kontizas}, {Kordopatis}, {Lanzafame}, {Lebreton}, {Licata}, {Lindstr{\o}m},
  {Livanou}, {Magdaleno Romeo}, {Manteiga}, {Marocco}, {Marshall}, {Mary},
  {Nicolas}, {Pallas-Quintela}, {Panem}, {Pichon}, {Poggio}, {Riclet}, {Robin},
  {Santove{\~n}a}, {Silvelo}, {Slezak}, {Smart}, {Soubiran}, {S{\"u}veges},
  {Ulla}, {Utrilla}, {Vallenari}, {Zhao}, {Zorec}, {Barrado}, {Bijaoui},
  {Bouret}, {Blomme}, {Brott}, {Cassisi}, {Kochukhov}, {Martayan}, {Shulyak},
  \& {Silvester}}]{Creevey2022}
{Creevey}, O.~L., {Sordo}, R., {Pailler}, F., {et~al.} 2022, arXiv e-prints,
  arXiv:2206.05864.
\newblock \doarXiv{2206.05864}

\bibitem[{Dierckx(1981)}]{Dierckx1981}
Dierckx, P. 1981, IMA Journal of Numerical Analysis, 1, 267,
  \dodoi{10.1093/imanum/1.3.267}

\bibitem[{Dotter {et~al.}(2008)Dotter, Chaboyer, Jevremovi{\'c}, Kostov, Baron,
  \& Ferguson}]{Dotter2008}
Dotter, A., Chaboyer, B., Jevremovi{\'c}, D., {et~al.} 2008, The Astrophysical
  Journal Supplement Series, 178, 89

\bibitem[{{Feiden} {et~al.}(2021){Feiden}, {Skidmore}, \& {Jao}}]{Feiden2021}
{Feiden}, G.~A., {Skidmore}, K., \& {Jao}, W.-C. 2021, \apj, 907, 53,
  \dodoi{10.3847/1538-4357/abcc03}

\bibitem[{{Ferguson} {et~al.}(2005){Ferguson}, {Alexander}, {Allard}, {Barman},
  {Bodnarik}, {Hauschildt}, {Heffner-Wong}, \& {Tamanai}}]{Ferguson2005}
{Ferguson}, J.~W., {Alexander}, D.~R., {Allard}, F., {et~al.} 2005, \apj, 623,
  585, \dodoi{10.1086/428642}

\bibitem[{{Fontes} {et~al.}(2015){Fontes}, {Zhang}, {Abdallah}, {Clark},
  {Kilcrease}, {Colgan}, {Cunningham}, {Hakel}, {Magee}, \&
  {Sherrill}}]{Fontes2016}
{Fontes}, C.~J., {Zhang}, H.~L., {Abdallah}, J., J., {et~al.} 2015, Journal of
  Physics B Atomic Molecular Physics, 48, 144014,
  \dodoi{10.1088/0953-4075/48/14/144014}

\bibitem[{{Gaia Collaboration} {et~al.}(2021){Gaia Collaboration}, {Smart},
  {Sarro}, {Rybizki}, {Reyl{\'e}}, {Robin}, {Hambly}, {Abbas}, {Barstow}, {de
  Bruijne}, \& et~al.}]{GaiaCollaboration2021}
{Gaia Collaboration}, {Smart}, R.~L., {Sarro}, L.~M., {et~al.} 2021, \aap, 649,
  A6, \dodoi{10.1051/0004-6361/202039498}

\bibitem[{{Grevesse} \& {Sauval}(1998)}]{Grevesse1998}
{Grevesse}, N., \& {Sauval}, A.~J. 1998, \ssr, 85, 161,
  \dodoi{10.1023/A:1005161325181}

\bibitem[{{Hakel} {et~al.}(2006){Hakel}, {Sherrill}, {Mazevet}, {Abdallah},
  {Colgan}, {Kilcrease}, {Magee}, {Fontes}, \& {Zhang}}]{Hakel2006}
{Hakel}, P., {Sherrill}, M.~E., {Mazevet}, S., {et~al.} 2006, \jqsrt, 99, 265,
  \dodoi{10.1016/j.jqsrt.2005.04.007}

\bibitem[{{Iglesias} \& {Rogers}(1996)}]{Iglesias1996}
{Iglesias}, C.~A., \& {Rogers}, F.~J. 1996, \apj, 464, 943,
  \dodoi{10.1086/177381}

\bibitem[{{Irwin}(2012)}]{Irwin2012}
{Irwin}, A.~W. 2012, {FreeEOS: Equation of State for stellar interiors
  calculations}, Astrophysics Source Code Library, record ascl:1211.002.
\newblock \doeprint{1211.002}

\bibitem[{{Jao} \& {Feiden}(2020)}]{Jao2020}
{Jao}, W.-C., \& {Feiden}, G.~A. 2020, \aj, 160, 102,
  \dodoi{10.3847/1538-3881/aba192}

\bibitem[{{Jao} \& {Feiden}(2021)}]{Jao2021}
---. 2021, Research Notes of the American Astronomical Society, 5, 124,
  \dodoi{10.3847/2515-5172/ac053a}

\bibitem[{{Jao} {et~al.}(2018){Jao}, {Henry}, {Gies}, \& {Hambly}}]{Jao2018}
{Jao}, W.-C., {Henry}, T.~J., {Gies}, D.~R., \& {Hambly}, N.~C. 2018, \apjl,
  861, L11, \dodoi{10.3847/2041-8213/aacdf6}

\bibitem[{{Jermyn} {et~al.}(2022){Jermyn}, {Bauer}, {Schwab}, {Farmer}, {Ball},
  {Bellinger}, {Dotter}, {Joyce}, {Marchant}, {Mombarg}, {Wolf}, {Wong},
  {Cinquegrana}, {Farrell}, {Smolec}, {Thoul}, {Cantiello}, {Herwig}, {Toloza},
  {Bildsten}, {Townsend}, \& {Timmes}}]{Jermyn2022}
{Jermyn}, A.~S., {Bauer}, E.~B., {Schwab}, J., {et~al.} 2022, arXiv e-prints,
  arXiv:2208.03651.
\newblock \doarXiv{2208.03651}

\bibitem[{{Joyce} \& {Chaboyer}(2018)}]{Joyce2018}
{Joyce}, M., \& {Chaboyer}, B. 2018, \apj, 864, 99,
  \dodoi{10.3847/1538-4357/aad464}

\bibitem[{{MacDonald} \& {Gizis}(2018)}]{MacDonald2018}
{MacDonald}, J., \& {Gizis}, J. 2018, \mnras, 480, 1711,
  \dodoi{10.1093/mnras/sty1888}

\bibitem[{{Magee} {et~al.}(2004){Magee}, {Abdallah}, {Colgan}, {Hakel},
  {Kilcrease}, {Mazevet}, {Sherrill}, {Fontes}, \& {Zhang}}]{Magee2004}
{Magee}, N.~H., {Abdallah}, J., {Colgan}, J., {et~al.} 2004, in American
  Institute of Physics Conference Series, Vol. 730, Atomic Processes in
  Plasmas: 14th APS Topical Conference on Atomic Processes in Plasmas, ed.
  J.~S. {Cohen}, D.~P. {Kilcrease}, \& S.~{Mazavet}, 168--179,
  \dodoi{10.1063/1.1824868}

\bibitem[{{Mansfield} \& {Kroupa}(2021)}]{Mansfield2021}
{Mansfield}, S., \& {Kroupa}, P. 2021, \aap, 650, A184,
  \dodoi{10.1051/0004-6361/202140536}

\bibitem[{{Nutzman} \& {Charbonneau}(2008)}]{Nut08}
{Nutzman}, P., \& {Charbonneau}, D. 2008, \pasp, 120, 317,
  \dodoi{10.1086/533420}

\bibitem[{Richardson(2007)}]{richardson2007beautiful}
Richardson, L. 2007, April

\bibitem[{{Rodr{\'\i}guez-L{\'o}pez}(2019)}]{Rodriguez-Lopez2019}
{Rodr{\'\i}guez-L{\'o}pez}, C. 2019, Frontiers in Astronomy and Space Sciences,
  6, 76, \dodoi{10.3389/fspas.2019.00076}

\bibitem[{{Seaton} {et~al.}(1994){Seaton}, {Yan}, {Mihalas}, \&
  {Pradhan}}]{Seaton1994}
{Seaton}, M.~J., {Yan}, Y., {Mihalas}, D., \& {Pradhan}, A.~K. 1994, \mnras,
  266, 805, \dodoi{10.1093/mnras/266.4.805}

\bibitem[{{Skrutskie} {et~al.}(2006){Skrutskie}, {Cutri}, {Stiening},
  {Weinberg}, {Schneider}, {Carpenter}, {Beichman}, {Capps}, {Chester},
  {Elias}, {Huchra}, {Liebert}, {Lonsdale}, {Monet}, {Price}, {Seitzer},
  {Jarrett}, {Kirkpatrick}, {Gizis}, {Howard}, {Evans}, {Fowler}, {Fullmer},
  {Hurt}, {Light}, {Kopan}, {Marsh}, {McCallon}, {Tam}, {Van Dyk}, \&
  {Wheelock}}]{Skrutskie2006}
{Skrutskie}, M.~F., {Cutri}, R.~M., {Stiening}, R., {et~al.} 2006, \aj, 131,
  1163, \dodoi{10.1086/498708}

\bibitem[{Sollima(2019)}]{Sollima2019}
Sollima, A. 2019, Monthly Notices of the Royal Astronomical Society, 489, 2377,
  \dodoi{10.1093/mnras/stz2093}

\bibitem[{{Spada} {et~al.}(2021){Spada}, {Demarque}, \& {Kupka}}]{Spada2021}
{Spada}, F., {Demarque}, P., \& {Kupka}, F. 2021, \mnras, 504, 3128,
  \dodoi{10.1093/mnras/stab1106}

\bibitem[{{Trampedach} {et~al.}(2014){Trampedach}, {Stein},
  {Christensen-Dalsgaard}, {Nordlund}, \& {Asplund}}]{Trampedach2014}
{Trampedach}, R., {Stein}, R.~F., {Christensen-Dalsgaard}, J., {Nordlund},
  {\r{A}}., \& {Asplund}, M. 2014, \mnras, 445, 4366,
  \dodoi{10.1093/mnras/stu2084}

\bibitem[{{van Saders} \& {Pinsonneault}(2012)}]{van2012}
{van Saders}, J.~L., \& {Pinsonneault}, M.~H. 2012, \apj, 751, 98,
  \dodoi{10.1088/0004-637X/751/2/98}

\bibitem[{Virtanen {et~al.}(2020)Virtanen, Gommers, Oliphant, Haberland, Reddy,
  Cournapeau, Burovski, Peterson, Weckesser, Bright, {van der Walt}, Brett,
  Wilson, Millman, Mayorov, Nelson, Jones, Kern, Larson, Carey, Polat, Feng,
  Moore, {VanderPlas}, Laxalde, Perktold, Cimrman, Henriksen, Quintero, Harris,
  Archibald, Ribeiro, Pedregosa, {van Mulbregt}, \& {SciPy 1.0
  Contributors}}]{2020SciPy-NMeth}
Virtanen, P., Gommers, R., Oliphant, T.~E., {et~al.} 2020, Nature Methods, 17,
  261, \dodoi{10.1038/s41592-019-0686-2}

\end{thebibliography}
\bibliographystyle{aasjournal}

\end{document}